\documentclass[11pt,a4paper]{article}
\pdfoutput=1
\usepackage{graphicx,epsfig}
\usepackage{dcolumn} 
\usepackage{slashed,color,amsmath,amssymb}
\usepackage{jheppub}
\usepackage{enumitem}
\usepackage{multirow}
\usepackage{mathrsfs} 
\usepackage{subcaption}

\usepackage{caption, subcaption}
\usepackage{multirow}
\usepackage{verbatim}
\newcommand{\be}{\begin{equation}}
\newcommand{\ee}{\end{equation}}
\newcommand{\bea}{\begin{eqnarray}}
\newcommand{\eea}{\end{eqnarray}}
\newcommand{\bit}{\begin{itemize}}
\newcommand{\eit}{\end{itemize}}

\def\gsim{\lower0.5ex\hbox{$\:\buildrel >\over\sim\:$}}
\def\lsim{\lower0.5ex\hbox{$\:\buildrel <\over\sim\:$}}

\hypersetup{
   colorlinks=true,       
   linkcolor=blue,        
   citecolor=red,         
   filecolor=magenta,     
}

\usepackage{bm}

\preprint{}

\title{$Z’$- Mediated Slepton and Sneutrino Signatures at  HL-LHC
} 

\author{Kareem Ezzat$^{1,2}$, Nidal Chamoun$^{3,4,5}$, Shaaban Khalil$^2$}

\affiliation{\vspace*{0.1in}$^1$ Department of Mathematics, Faculty of Science, Ain Shams University, Cairo 11566, Egypt.}
\affiliation{\vspace*{0.1in}$^2$ Center for Fundamental Physics, Zewail City of Science and
	Technology, 6th of October City, Giza 12578, Egypt}
\affiliation{\vspace*{0.1in}$^3$ Physics Department, HIAST, P.O. Box 31983, Damascus, Syria.}
\affiliation{\vspace*{0.1in}$^4$ CASP, Antioch Syrian University, Maaret Saidnaya, Damascus, Syria.}
\affiliation{\vspace*{0.1in}$^5$ Bethe Center for Theoretical Physics and Physikalisches Institut der Universit\"at Bonn, \\
	Nußallee 12, 53115 Bonn, Germany}

\emailAdd{kareemezat@sci.asu.edu.eg}
\emailAdd{chamoun@uni-bonn.de}
\emailAdd{skhalil@zewailcity.edu.eg}

\abstract{ We investigate the discovery potential of a heavy $Z'$ boson as a portal to charged sleptons and right-handed sneutrinos within the $B-L$ Supersymmetric Standard Model (BLSSM) at the High-Luminosity LHC. Using a detector-level simulation and Boosted Decision Trees to suppress Standard Model backgrounds, we find that the direct slepton decay $pp \to Z' \to \tilde{\ell}^+\tilde{\ell}^- \to 2\ell + E_{\rm T}^{\rm miss}$ yields a significance exceeding $5\sigma$ at 3000 fb$^{-1}$, while the cascade $(\mbox{multileptons} + E_{\rm T}^{\rm miss})$ mode is unobservable due to rate suppression. For right-handed sneutrinos, the hadronic channel $pp \to Z' \to \tilde{\nu}_R\tilde{\nu}_I \to 2\ell + 4j + E{\rm T}^{\rm miss}$ emerges as the most promising signature. These results establish the $2\ell + E_{\rm T}^{\rm miss}$ and $2\ell + 4j + E_{\rm T}^{\rm miss}$ final states as the principal discovery channels for the scalar leptonic sector of the BLSSM at the HL-LHC.}

\begin{document}

	\maketitle

\section{Introduction}
\label{sec:intro}

The $B-L$ Supersymmetric Standard Model (BLSSM) provides a
well-motivated extension of the Minimal Supersymmetric Standard Model
(MSSM)\cite{KHALIL2008374,OLeary:2011yq,Basso:2011na,Basso:2015pka}, in which the SM gauge group is enlarged by an additional
$U(1)_{B-L}$ symmetry. This extension provides a natural framework for
neutrino mass generation through the introduction of right-handed
neutrinos. The additional gauge symmetry is associated with a new
neutral gauge boson, $Z'$, whose mass is generated through spontaneous
$B-L$ symmetry breaking. Its interactions with both SM and
supersymmetric particles lead to a rich collider phenomenology and
provide new production mechanisms for states that may otherwise be
difficult to probe at hadron colliders.

An important consequence of supersymmetrizing the $B-L$ framework is
that the right-handed neutrinos are accompanied by their scalar
superpartners, the right-handed sneutrinos. The BLSSM therefore contains,
in addition to the MSSM charged sleptons, an extended sneutrino sector
that can be directly accessed through the $Z'$ gauge interaction.
In particular, the resonant processes
\begin{equation}
pp\to Z'\to\tilde{\ell}^{+}\tilde{\ell}^{-},
\qquad
pp\to Z'\to\tilde{\nu}_R \tilde{\nu}_I,
\label{eq:intro_scalar_channels}
\end{equation}
provide an alternative to the conventional electroweak production of
sleptons and sneutrinos and can substantially enhance their production
rates when the $Z'$ is produced on shell.

The collider phenomenology of the $Z'$ itself has been extensively
investigated through its decays into SM dileptons, heavy neutrinos,
Higgs bosons, and supersymmetric states. In particular, the
$Z'$-mediated production of heavy right-handed neutrinos was recently
studied in Ref.~\cite{Ezzat:2024twd} through
\begin{equation}
pp\to Z'\to N_R N_R,
\end{equation}
followed by $N_R\to\ell W$. Depending on the subsequent $W$-boson
decays, three characteristic signatures were analyzed:
$2\ell+4j$, $4\ell+E_T^{\rm miss}$, and
$3\ell+2j+E_T^{\rm miss}$. A multivariate analysis demonstrated
significant discovery potential for all three channels at the {High-Luminosity Large Hadron Collider} (HL-LHC),
showing that resonant $Z'$ production provides an efficient probe of
the heavy right-handed-neutrino sector.

The supersymmetric BLSSM opens qualitatively different possibilities. Actually, previous collider studies have investigated right-handed sneutrino
phenomenology in BLSSM, including multilepton and missing-energy signatures
~\cite{DelleRose:2017ukx,Huitu:2008gf,Elsayed:2012ec,
	Arina:2013zca,Bandyopadhyay:2011es,Abdallah:2015uba}.
Resonant charged-slepton production through an additional $Z'$ boson
has also been explored in BLSSM and in other supersymmetric extensions involving different $U(1)$
assignments~\cite{Corcella:2014xwa,Baumgart:2006pa,Krauss:2012ku}.
	
	Building upon these studies, we perform a dedicated detector-level
	analysis of $Z'$-mediated charged-slepton and right-handed-sneutrino
	production at the HL-LHC. A central feature of the present analysis is
	the use of multivariate techniques based on boosted decision trees
	(BDTs)~\cite{Xia:2018cfz,XIA201915,quinlan1986induction,
		quinlan1987simplifying}, implemented within the TMVA
	framework~\cite{hoecker2007tmva}. Rather than relying solely on
	conventional cut-based selections, the BDT combines a broad set of
	kinematic and angular observables and exploits correlations among them
	to enhance the discrimination between the signal and the SM backgrounds. This strategy enables a systematic assessment of
	the HL-LHC sensitivity to several direct and cascade decay channels of
	charged sleptons and right-handed sneutrinos within a common analysis
	framework.

In this work, we focus on the largely unexplored resonant production of
charged sleptons through
$Z'\to\tilde{\ell}^{+}\tilde{\ell}^{-}$ and investigate both their
direct and cascade decays. In particular, the direct decay
$\tilde{\ell}^{\pm}\to\ell^{\pm}\tilde{\chi}_1^0$ leads to the
$2\ell+E_T^{\rm miss}$ signature. Although this final state suffers
from sizeable SM backgrounds, the resonant enhancement of the slepton
production rate through the $Z'$ makes it a promising channel at the
HL-LHC. We also examine multilepton signatures arising from cascade
slepton decays, where the output includes $4$ or $6$ leptons depending on whether one or both branches undergo cascade decay, and assess whether their cleaner background environment
can compensate for the additional branching-ratio suppression.

We further investigate the resonant production of right-handed
sneutrinos through
$Z'\to\tilde{\nu}_R \tilde{\nu}_I$. Unlike the fermionic
right-handed neutrinos considered in Ref.~\cite{Ezzat:2024twd}, the
right-handed sneutrinos undergo supersymmetric cascade decays involving
charginos and neutralinos. Their collider signatures therefore carry
missing transverse momentum from the stable lightest neutralinos and
provide a distinct probe of the scalar sector associated with
right-handed neutrinos. We consider fully leptonic, semileptonic, and
hadronic $W$-decay modes and determine which of these channels can be
realistically probed at the HL-LHC.

The aim of this work is therefore twofold: to assess the discovery
potential of $Z'$-mediated charged-slepton production, with particular
emphasis on the direct $2\ell+E_T^{\rm miss}$ channel, and to contrast it with the 
 HL-LHC sensitivity to the separate channel of right-handed sneutrinos. Our
analysis demonstrates that the resonant $Z'$ portal can considerably
improve the prospects for probing the scalar leptonic sector of the
BLSSM, while the different decay patterns of sleptons, right-handed
sneutrinos, and right-handed neutrinos provide complementary handles for
disentangling the underlying $B-L$ dynamics.

The remainder of this paper is organized as follows.
In Sec.~\ref{sec:BLSSM}, we briefly review the main features of the
BLSSM relevant to the present study, with particular emphasis on the
benchmark scenarios adopted in our analysis.
In Sec.~\ref{sec:Zdecays}, we discuss the $Z'$ decays into charged
sleptons and right-handed sneutrinos and identify the corresponding
LHC signal topologies.
Section~\ref{sec:collider} is devoted to the collider analysis at the
HL-LHC. We first investigate the direct slepton decay channel,
$2\ell+E_T^{\rm miss}$, followed by { cascade(s)}  slepton signature,
${4(6)}\ell+E_T^{\rm miss}$, and then analyze the {complementary}
right-handed sneutrino channels.
Finally, our conclusions and closing remarks are presented in
Sec.~\ref{sec:conclusion}.

\section{The \boldmath{$B-L$} Supersymmetric Standard Model}
\label{sec:BLSSM}

The $B-L$ Supersymmetric Standard Model (BLSSM) extends the gauge
symmetry of the Minimal Supersymmetric Standard Model (MSSM) according to
\begin{equation}
SU(3)_C \times SU(2)_L \times U(1)_Y
\times U(1)_{B-L}.
\label{eq:BLSSMgauge}
\end{equation}
Anomaly cancellation requires the introduction of three generations of
right-handed neutrino superfields, $\widehat N_i^c$. In addition, two
SM singlet Higgs superfields, $\widehat\chi_1$ and
$\widehat\chi_2$, carrying opposite $B-L$ charges, are introduced to
spontaneously break the additional Abelian symmetry. Their vacuum
expectation values generate the mass of the new neutral gauge boson
$Z'$. Since the BLSSM has been extensively discussed in the literature,
we summarize here only the ingredients relevant to the slepton and
sneutrino collider signatures studied in this work.

The gauge interactions are described by the covariant derivative
\begin{equation}
D_\mu
=
\partial_\mu
-i g_s T^\alpha G_\mu^\alpha
-i g_2 \tau^a W_\mu^a
-i g_1 Y B_\mu
-i\left(\widetilde g\,Y+g_{BL}Y_{B-L}\right)B'_\mu ,
\label{eq:BLSSMcovariant}
\end{equation}
where $g_s$, $g_2$, $g_1$, and $g_{BL}$ are the gauge couplings
associated with $SU(3)_C$, $SU(2)_L$, $U(1)_Y$, and $U(1)_{B-L}$,
respectively. The parameter $\widetilde g$ denotes the gauge kinetic
mixing between the two Abelian gauge groups.

The renormalizable superpotential is given by
\begin{align}
W ={}&
Y_u^{ij}\,\widehat u_i^c\widehat Q_j\!\cdot\!\widehat H_u
-Y_d^{ij}\,\widehat d_i^c\widehat Q_j\!\cdot\!\widehat H_d
-Y_e^{ij}\,\widehat e_i^c\widehat L_j\!\cdot\!\widehat H_d
\nonumber\\
&+
Y_\nu^{ij}\,\widehat N_i^c\widehat L_j\!\cdot\!\widehat H_u
+\frac{1}{2}Y_N^{ij}\,
 \widehat N_i^c\widehat\chi_1\widehat N_j^c
+\mu\,\widehat H_u\!\cdot\!\widehat H_d
-\mu'\widehat\chi_1\widehat\chi_2 .
\label{eq:BLSSMsuperpotential}
\end{align}
The first three terms reproduce the MSSM Yukawa interactions, while
$Y_\nu$ and $Y_N$ determine the neutrino and right-handed sneutrino
sectors.

After the neutral scalar fields acquire vacuum expectation values,
\begin{equation}
\left\langle H_u^0\right\rangle=\frac{v_u}{\sqrt{2}},
\qquad
\left\langle H_d^0\right\rangle=\frac{v_d}{\sqrt{2}},
\qquad
\left\langle\chi_1\right\rangle=\frac{v_1}{\sqrt{2}},
\qquad
\left\langle\chi_2\right\rangle=\frac{v_2}{\sqrt{2}},
\end{equation}
we define
\begin{equation}
v=\sqrt{v_u^2+v_d^2},
\qquad
v'=\sqrt{v_1^2+v_2^2},
\qquad
\tan\beta=\frac{v_u}{v_d},
\qquad
\tan\beta'=\frac{v_1}{v_2}.
\end{equation}
The $Z'$ mass is then approximately
\begin{equation}
M_{Z'}^2
=
g_{BL}^2v'^2+\frac{1}{4}\widetilde g^{\,2}v^2.
\label{eq:ZprimeMass}
\end{equation}
The mixing between the SM-like $Z$ boson and the $Z'${, expressed in terms of the parameter $\theta'$}, is strongly
constrained by electroweak precision measurements{, and thus we} work
in the small-mixing limit,
\begin{equation}
|\theta'|\lesssim 10^{-3}.
\end{equation}

\subsection{Benchmark scenarios}

To study the two main signal topologies, we introduce two benchmark
scenarios. We first consider a charged-slepton benchmark, denoted by
BP-I, for the resonant process
\begin{equation}
pp\to Z'\to\widetilde\ell_1^+\widetilde\ell_1^- .
\label{eq:BPslepton}
\end{equation}
The lightest charged slepton must satisfy
\begin{equation}
2m_{\widetilde\ell_1}<M_{Z'}
\end{equation}
and is chosen to decay predominantly through
\begin{equation}
\widetilde\ell_1^\pm
\to \ell^\pm\widetilde\chi_1^0 .
\end{equation}
The sneutrino states are taken sufficiently heavy to suppress competing
$Z'$ decays into sneutrino pairs and thereby enhance the charged-slepton
branching fraction.

The second benchmark, BP-II, is designed for resonant production of the
lightest CP-even and CP-odd predominantly right-handed sneutrinos,
\begin{equation}
pp\to Z'\to\widetilde\nu_{R_1}\widetilde\nu_{I_1}.
\label{eq:BPsneutrino}
\end{equation}
In this case, the sneutrino states are light enough to be produced
on shell, while the charged sleptons are heavier. The dominant
sneutrino decay chain proceeds through charginos and produces the
multilepton signatures discussed in the following section.

The gauge-sector parameters common to the two benchmarks are summarized
in Table~\ref{tab:input}. The corresponding relevant mass spectra are
given in Table~\ref{tab:spectrum}.
%
\begin{table}[t]
\centering
\renewcommand{\arraystretch}{1.2}
\begin{tabular}{cccccc}
\hline\hline
$g_{BL}$ & $\widetilde g$ & $\tan\beta$ & $\tan\beta'$
& $v'$ & $(\mu,\mu')$ \\
\hline
$0.5$ & $-0.6$ & $10$ & $1.2$
& $5441$ & $(1676,\,914)$ \\
\hline\hline
\end{tabular}
\caption{Common input parameters used for the two benchmark scenarios.
The dimensionful parameters $v'$, $\mu$, and $\mu'$ are given in GeV.}
\label{tab:input}
\end{table}
%
\begin{table}[t]
\centering
\renewcommand{\arraystretch}{1.2}
\begin{tabular}{ccc}
\hline\hline
Particle
& BP-I: $\widetilde\ell_1^+\widetilde\ell_1^-$ channel
& BP-II: $\widetilde\nu_{R_1}\widetilde\nu_{I_1}$ channel \\
\hline
$M_{Z'}$                         & $3000$ & $3000$ \\
$m_{h_1}$                        & $123$     & $123$  \\
$m_{\widetilde\chi_1^0}$         & $520$     & $661$  \\
$m_{\widetilde\chi_2^0}$         & $650$     & $763$  \\
$m_{\widetilde\chi_1^\pm}$       & $650$     & $763$  \\
$m_{\widetilde\ell_1}$           & $2000$     & $820$ \\
$m_{\widetilde\nu_{R_1}}$        & $700$     & $1229$ \\
$m_{\widetilde\nu_{I_1}}$        & $690$     & $1209$ \\
\hline\hline
\end{tabular}
\caption{Relevant mass spectra for the charged-slepton benchmark BP-I
and the right-handed sneutrino benchmark BP-II. All masses are given
in GeV. }
\label{tab:spectrum}
\end{table}

%
For the considered benchmark parameters, the effective coupling of the
$Z'$ boson to a fermion $f$ is controlled by ~\cite{Basso:2009hf,Chun:2017spz}:
\begin{equation}
g_f^{Z'}
=
g_{BL}Q_{B-L}^{f}
+\widetilde g\,Y_f ,
\label{eq:effectiveZcoupling}
\end{equation}
so that for charged leptons, $Q_{B-L}^{\ell}=-1$ and $Y_\ell=-1$, and the considered
benchmark values, one gets
\begin{equation}
g_{\ell}^{Z'}
=
-0.5+0.6
\simeq
0.1.
\label{eq:effectiveZlepton}
\end{equation}
The resulting effective leptonic coupling is therefore relatively small,
leading to a suppressed Drell--Yan dilepton branching fraction of the
$Z'$ boson. In addition, the opening of decay channels into supersymmetric
particles further reduces
\begin{equation}
\operatorname{BR}\!\left(Z'\to\ell^+\ell^-\right).
\end{equation}
Consequently, the benchmark value $M_{Z'}=3~\mathrm{TeV}$ remains
phenomenologically viable within the assumptions of the present analysis
and motivates searches based on resonant $Z'$ decays into sleptons and
sneutrinos.

\section{\boldmath{$Z'$} Decays into Sleptons and Sneutrinos at the LHC}
\label{sec:Zdecays}

In addition to the conventional dilepton channel,
\begin{equation}
pp \to Z' \to \ell^+\ell^-,
\end{equation}
a heavy neutral gauge boson can also provide an efficient portal to the
supersymmetric sector if kinematically allowed. In particular, the
resonant production of sleptons and sneutrinos through an intermediate
$Z'$ boson can lead to distinctive multi-lepton signatures at the LHC.
Unlike the standard Drell--Yan production of leptons, the corresponding
production cross sections can be significantly enhanced near the $Z'$
resonance, thereby extending the sensitivity to heavier supersymmetric
states.

In this work, we focus on the processes
\begin{equation}
pp \to Z' \to \tilde{\ell}^+ \tilde{\ell}^- ,
\end{equation}
and
\begin{equation}
pp \to Z' \to \tilde{\nu}_R \tilde{\nu}_I,
\end{equation}
where $\tilde{\ell}$ denotes charged sleptons and $\tilde{\nu}_{I,R}$
represent the real and imaginary parts of the lightest right-handed sneutrino. The subsequent decays of these
particles may generate final states containing charged leptons, jets,
and large missing transverse energy, leading to potentially clean and
observable collider signatures.

\subsection{Charged Slepton Channels}

If kinematically allowed, the heavy $Z'$ boson can decay into a pair of
charged sleptons,
\begin{equation}
pp\rightarrow Z'\rightarrow
\tilde{\ell}^+\tilde{\ell}^-.
\label{eq:sleptonproduction}
\end{equation}
Compared with the conventional Drell--Yan production mechanism, resonant
production through the $Z'$ can considerably enhance the production rate,
thereby extending the LHC sensitivity to heavier sleptons.

The simplest decay channel is
\begin{equation}
\tilde{\ell}^{\pm}
\rightarrow
\ell^{\pm}\tilde{\chi}_1^0,
\end{equation}
leading to the signature
\begin{equation}
\ell^+\ell^-+E_T^{\rm miss},
\end{equation}
which suffers from sizeable SM backgrounds, mainly from
$WW$, $t\bar t$, $ZZ$, and
$Z/\gamma^\ast\rightarrow\tau^+\tau^-$.

A cleaner signature is obtained through cascade decays,
\begin{equation}
\tilde{\ell}^{\pm}
\rightarrow
\ell^{\pm}\tilde{\chi}_2^0,
\end{equation}
followed by
\begin{equation}
\tilde{\chi}_2^0
\rightarrow
\ell^+\ell^-\tilde{\chi}_1^0,
\end{equation}
which gives rise{, for one (two) cascade(s),}  to
\begin{equation}
{4(6)}\ell+E_T^{\rm miss}.
\end{equation}
The multi-lepton topology provides a much cleaner probe of resonant
slepton production because of its substantially reduced SM
background.

\subsection{Right-Handed Sneutrino Channels}

The second class of signals arises from resonant production of
right-handed sneutrinos,
\begin{equation}
pp\rightarrow Z'
\rightarrow
\tilde{\nu}_R\tilde{\nu}_I .
\end{equation}
The dominant decay mode considered in this work is
\begin{equation}
\tilde{\nu}_{R,I}
\rightarrow
\ell^\pm\tilde{\chi}_1^\mp ,
\end{equation}
followed by
\begin{equation}
\tilde{\chi}_1^\pm
\rightarrow
W^\pm\tilde{\chi}_1^0,
\end{equation}
leading to the characteristic final state
\begin{equation}
\ell^+\ell^-+W^+W^-+E_T^{\rm miss}.
\end{equation}
Depending on the subsequent $W$-boson decays, three representative
final states are considered. Fully leptonic, semi-leptonic, and fully
hadronic $W^+W^-$ decays lead, respectively, to
\begin{equation}
	4\ell+E_T^{\rm miss},\qquad
	3\ell+2j+E_T^{\rm miss},\qquad
	2\ell+4j+E_T^{\rm miss}.
\end{equation}
Among these, the third predominantly hadronic mode benefits from the larger hadronic
branching fraction of the $W$ boson and provides the most promising
right-handed-sneutrino channel in our analysis.

Invisible decays,
\begin{equation}
\tilde{\nu}_{R,I}
\rightarrow
\nu\tilde{\chi}_1^0,
\end{equation}
lead to mono-jet or mono-photon signatures in association with
initial-state radiation.

\subsection{{Signal topologies}}

The above discussion identifies four representative signal topologies,
which form the basis of the collider analysis presented in the next section:
\begin{itemize}
\item[(i)] $2\ell+E_T^{\rm miss}$ from direct slepton decays;
\item[(ii)] ${4(6)}\ell+E_T^{\rm miss}$ from slepton {one (two) cascade(s)} decays;
\item[(iii)] $\ell^+\ell^-+W^+W^-+E_T^{\rm miss}$ from right-handed sneutrino decays; which in turn is divided
into subchannels depending on the nature of the subsequent W-decays, with the most significant being the subchannel $2\ell+4j+E_T^{\rm miss}$; 
\item[(iv)] Invisible or semi-visible sneutrino decays leading to mono-jet, mono-photon, or multi-lepton signatures.
\end{itemize}

\section{Collider Analysis}
\label{sec:collider}

In this section, we investigate the discovery prospects of the
$Z'$-mediated slepton and sneutrino signatures introduced in the previous
section at the HL-LHC. The
analysis is performed for proton--proton collisions at a center-of-mass
energy of $\sqrt{s}=14~\mathrm{TeV}$, assuming an integrated luminosity
of $\mathcal{L}=3000~\mathrm{fb}^{-1}$.

Following the signal topologies discussed in Section~3, we first analyze
the direct production of charged sleptons through the
$2\ell+E_T^{\rm miss}$ final state. We then investigate the {one (two) cascade(s)}
decay of charged sleptons leading to the
${4(6)}\ell+E_T^{\rm miss}$ signature. Finally, we consider the
right-handed sneutrino channels, including the fully leptonic,
semi-leptonic, and fully hadronic { $W$-decay} final states,
which provide complementary probes of the BLSSM parameter space.

For each signal topology, the dominant SM backgrounds are
identified and simulated. A common analysis strategy is adopted,
consisting of an initial event preselection followed by a multivariate
classification based on a Boosted Decision Tree (BDT). The BDT exploits
kinematic observables related to the visible decay products, missing
transverse momentum, and global event activity to maximize the separation
between signal and background events. The discovery potential of each
channel is finally quantified in terms of the expected statistical
significance at the HL-LHC.

\subsection{Event Simulation and Analysis Strategy}
%
Signal and background events are generated at leading order using
{\tt MadGraph5\_aMC@NLO}~\cite{Alwall:2014hca}, employing the BLSSM model implemented through
{\tt SARAH} package for Mathematica ~\cite{Staub:2013tta} and {\tt SPheno}~\cite{Porod:2003um,Porod:2011nf}. The generated parton-level events are subsequently interfaced to {\tt Pythia~8}~\cite{Sjostrand:2014zea} for parton showering,
hadronization, and particle decays. Detector effects are simulated using
{\tt Delphes}~\cite{deFavereau:2013fsa}, adopting the HL-LHC detector configuration.

Jets are reconstructed with the anti-$k_T$ clustering algorithm with a
distance parameter $R=0.4$ as implemented in {\tt FastJet}~\cite{Cacciari:2008gp}. Electrons
and muons are required to satisfy standard isolation criteria and are
selected within the detector acceptance. Identified $b$-jets are used
to suppress backgrounds originating from top-quark production whenever
appropriate.

For each signal topology, dedicated Monte Carlo samples are generated
for both the signal and the relevant { SM} backgrounds. After
applying the basic event preselection, a BDT
classifier is trained using a set of discriminating kinematic variables.
The performance of the classifier is evaluated using the receiver
operating characteristic (ROC) curve, while possible overtraining is
checked through the Kolmogorov--Smirnov (KS) test by comparing the
training and testing samples.

The final signal sensitivity is determined by optimizing the BDT output
selection and evaluating the expected Asimov significance
$Z_A$~\cite{Cowan:2010js}. To account for systematic uncertainties in
the background prediction, we use
\begin{equation}
	Z_A =
	\left\{
	2\left[
	(S+B)\ln\left(
	\frac{(S+B)(B+\sigma_B^2)}
	{B^2+(S+B)\sigma_B^2}
	\right)
	-
	\frac{B^2}{\sigma_B^2}
	\ln\left(
	1+
	\frac{\sigma_B^2 S}
	{B(B+\sigma_B^2)}
	\right)
	\right]
	\right\}^{1/2},
	\label{eq:Asimov_withSys}
\end{equation}
where $S$ and $B$ denote the expected signal and background yields,
respectively, and $\sigma_B$ represents the absolute uncertainty on the
background yield, parameterized as
\begin{equation}
	\sigma_B^2=\delta_B B,
\end{equation}
with $\delta_B$ denoting the fractional systematic uncertainty. We consider representative values of\footnote{For $\delta_B=0$, Eq. (\ref{eq:Asimov_withSys}) reduces to the simple form $Z_A =
		\sqrt{2\left[(S+B)\ln\left(1+\frac{S}{B}\right)-S\right]},$}
\begin{equation}
	\delta_B= 0\%,  5\%,\;10\%,\;20\%.
\end{equation}

\subsection{Direct Slepton Decays:
$\mathbf{2\ell+E_T^{\rm miss}}$}

We first investigate the direct production of the lightest charged
slepton through the resonant decay of the heavy neutral gauge boson,
\begin{equation}
pp
\rightarrow
Z'
\rightarrow
\tilde{\ell}_1^+
\tilde{\ell}_1^-,
\end{equation}
followed by the leptonic decay
\begin{equation}
\tilde{\ell}_1^\pm
\rightarrow
\ell^\pm
\tilde{\chi}_1^0,
\end{equation}
which leads to the experimentally clean signature
\begin{equation}
pp
\rightarrow
Z'
\rightarrow
\tilde{\ell}_1^+
\tilde{\ell}_1^-
\rightarrow
2\ell
+
E_T^{\rm miss},
\end{equation}
where $\ell=e,\mu$, and the missing transverse momentum originates from
the two invisible neutralinos.

Compared with the conventional Drell--Yan production mechanism,
resonant production through the heavy $Z'$ boson considerably enhances
the charged-slepton production cross section, thereby extending the
discovery reach for sleptons at the HL-LHC.




The dominant SM backgrounds are
\[
WW,\;
WWW,\;
WZ,\;
t\bar t,\;
tW,\;
Z/\gamma^\ast+\mathrm{jets},\;
ZZ\rightarrow2\ell2\nu,
\;
Z/\gamma^{*}\rightarrow\ell^{+}\ell^{-},
\]
among which the $WW$ and $t\bar t$ processes constitute the largest
backgrounds after the basic event selection. {At generator level, events are required to satisfy $m_{\ell\ell}>300~\mathrm{GeV},$ in order to suppress the low-mass Drell--Yan contribution and focus on
the high-mass region relevant to the heavy $Z'$ signal.}

Events are required to contain
\begin{itemize}
\item exactly two isolated opposite-sign leptons;
\item
$p_T(\ell_1)>25~\mathrm{GeV}$,
$p_T(\ell_2)>20~\mathrm{GeV}$,
with
$|\eta(\ell)|<2.5$;
\item
missing transverse energy
$E_T^{\rm miss}>50~\mathrm{GeV}$;
\item
no tagged $b$-jets. 
\end{itemize}  
These requirements efficiently suppress the dominant backgrounds,
particularly those originating from top-quark production, while
maintaining a high signal acceptance.

Table~\ref{cutflow_2lmet} summarizes the production cross sections,
generated Monte Carlo samples, surviving events after the basic
preselection and the optimized BDT selection, together with the
corresponding expected event yields for an integrated luminosity of
$3000~\mathrm{fb}^{-1}$.

\begin{table}[htbp]
	\centering
	\renewcommand{\arraystretch}{1.15}
	\begin{tabular}{lccccc}
		\hline
		Process &
		$\sigma$ (fb) &
		Generated &
		After preselection &
		After BDT &
		Expected events \\
		\hline
		Signal & 3.3062 & 400000 & 279816 & 153817 & 3814.53 \\
		$Z/\gamma^{*}\rightarrow\ell^{+}\ell^{-}$
		& 754.16 & 400000 & 4971 & 3 & 16.97 \\
		$ZZ\rightarrow2\ell2\nu$
		& 249.52 & 297493 & 50859 & 11 & 27.68 \\
		$WW\rightarrow2\ell2\nu$
		& 3360.487 & 400000 & 38756 & 13 & 327.65 \\
		$\ell^{+}\ell^{-}+j$
		& 421.2217 & 331447 & 8513 & 15 & 57.19 \\
		$t\bar{t}\rightarrow2\ell+2b+E_T^{\rm miss}$
		& 26893.6 & 400000 & 15483 & 4 & 806.81 \\
		$\ell^{+}\ell^{-}+jj$
		& 195.4846 & 400000 & 17370 & 24 & 35.19 \\
		$WWW\rightarrow3\ell+E_T^{\rm miss}$
		& 0.8437 & 400000 & 63482 & 553 & 3.50 \\
		$WZ\rightarrow3\ell+E_T^{\rm miss}$
		& 247.4589 & 400000 & 37184 & 7 & 12.99 \\
		\hline
		Total background & --- & --- & --- & --- & 1287.97 \\
		\hline
	\end{tabular}
	\caption{Production cross sections, generated events, surviving Monte Carlo
		events after the basic preselection and the optimized BDT selection, together
		with the corresponding expected event yields at the HL-LHC with an integrated
		luminosity of $3000~\mathrm{fb}^{-1}$.}
    \label{cutflow_2lmet}
\end{table}


The basic event selection suppresses the SM backgrounds by
more than two orders of magnitude while preserving approximately
70\% of the signal events.
The optimized BDT selection further improves the signal purity,
reducing the total background to only
$1288$
expected events while retaining approximately
55\%
{of the preselected signal events.}

{
	To maximize the separation between signal and background, a BDT classifier is trained using a set of kinematic
	observables characterizing the two leptons, the missing transverse
	momentum, and the overall event topology. The input variables include
	$p_T(\ell_1)$, $p_T(\ell_2)$, $m_{\ell\ell}$, $E_T^{\rm miss}$,
	$H_T$, $L_T$, $S_T$, $m_T(\ell_{1,2},E_T^{\rm miss})$, $M_{T2}$,
	$p_T(\ell\ell)$, several angular separations, the ratios
	$E_T^{\rm miss}/S_T$ and $E_T^{\rm miss}/L_T$, and the jet multiplicity.
	Here, $H_T$ denotes the sum of the transverse momenta of the
	selected jets, while $L_T=p_T(\ell_1)+p_T(\ell_2)$ represents the
	leptonic transverse activity, and
	$S_T=H_T+L_T+E_T^{\rm miss}$ characterizes the total transverse
	activity of the event. The stransverse mass $M_{T2}$ is particularly
	useful for pair-production topologies involving two invisible
	particles, as in the slepton decay considered here.

 The ranking of the ten most discriminating observables is summarized in
Table~\ref{tab:ranking_2lMET}. Among these,
$E_T^{\rm miss}/S_T$,
$m_{\ell\ell}$,
and
$m_T(\ell_1,E_T^{\rm miss})$
are found to provide the strongest discrimination between the signal
and the SM backgrounds.}

\begin{table}[htbp]
	\centering
	\begin{tabular}{lcc}
		\hline
		Rank & Variable & Importance \\
		\hline
		1  & $E_T^{\rm miss}/S_T$                    & 0.07526 \\
		2  & $m_{\ell\ell}$                          & 0.07029 \\
		3  & $m_T(\ell_1,E_T^{\rm miss})$           & 0.06950 \\
		4  & $p_T(\ell_2)$                           & 0.06471 \\
		5  & $\Delta R_{\ell\ell}$                   & 0.06000 \\
		6  & $\Delta\phi(\ell_2,E_T^{\rm miss})$     & 0.05752 \\
		7  & $E_T^{\rm miss}$                        & 0.05584 \\
		8  & $m_T(\ell_2,E_T^{\rm miss})$           & 0.05501 \\
		9  & $\Delta\phi(\ell\ell,E_T^{\rm miss})$   & 0.05462 \\
		10 & $\Delta\phi(\ell_1,\ell_2)$             & 0.05418 \\
		\hline
	\end{tabular}
	\caption{Ranking of the ten most important BDT input variables for the
		$2\ell+E_T^{\rm miss}$ final state.}
	\label{tab:ranking_2lMET}
\end{table}

\begin{figure}[htbp]
	\centering
	
	\begin{subfigure}{0.32\textwidth}
		\centering
		\includegraphics[width=\linewidth]{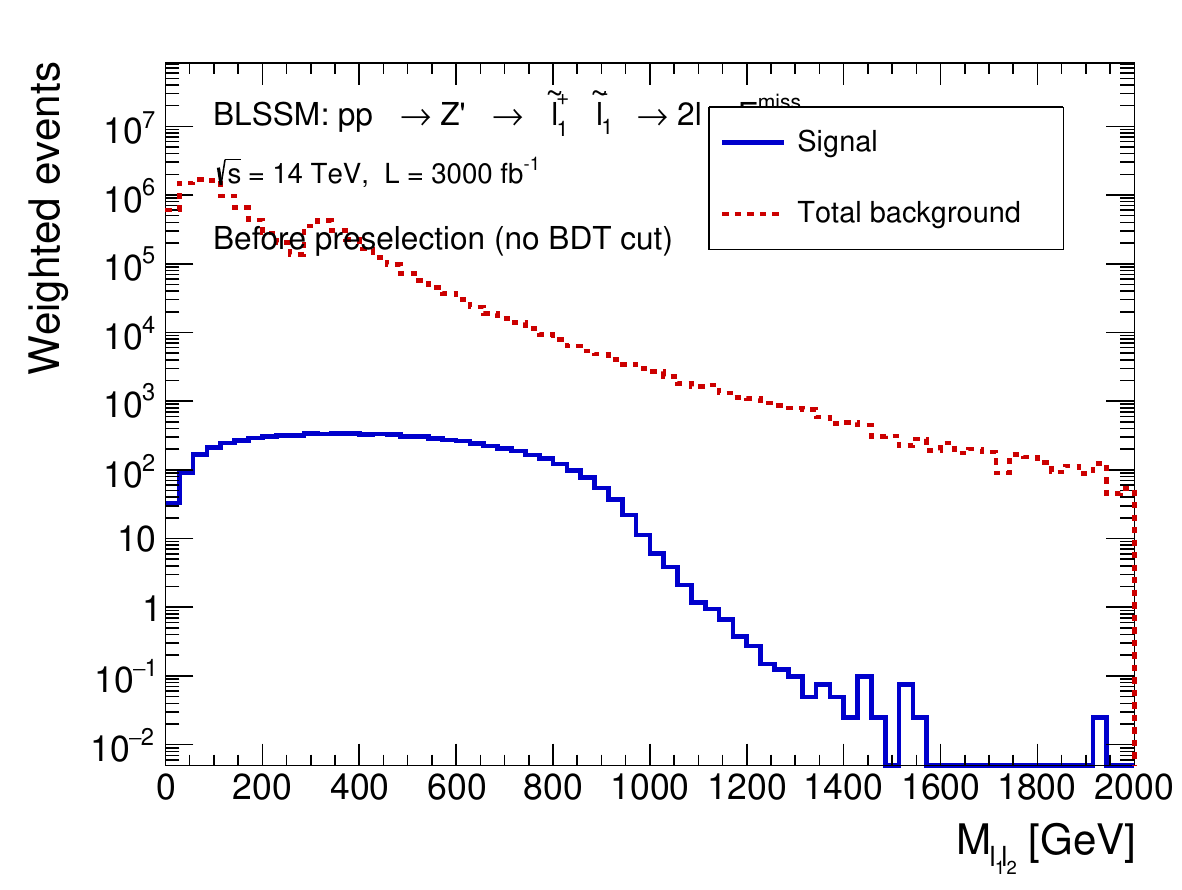}
		\caption{Before preselection.}
		\label{fig:mll_before_2lMET}
	\end{subfigure}
	\hfill
	\begin{subfigure}{0.32\textwidth}
		\centering
		\includegraphics[width=\linewidth]{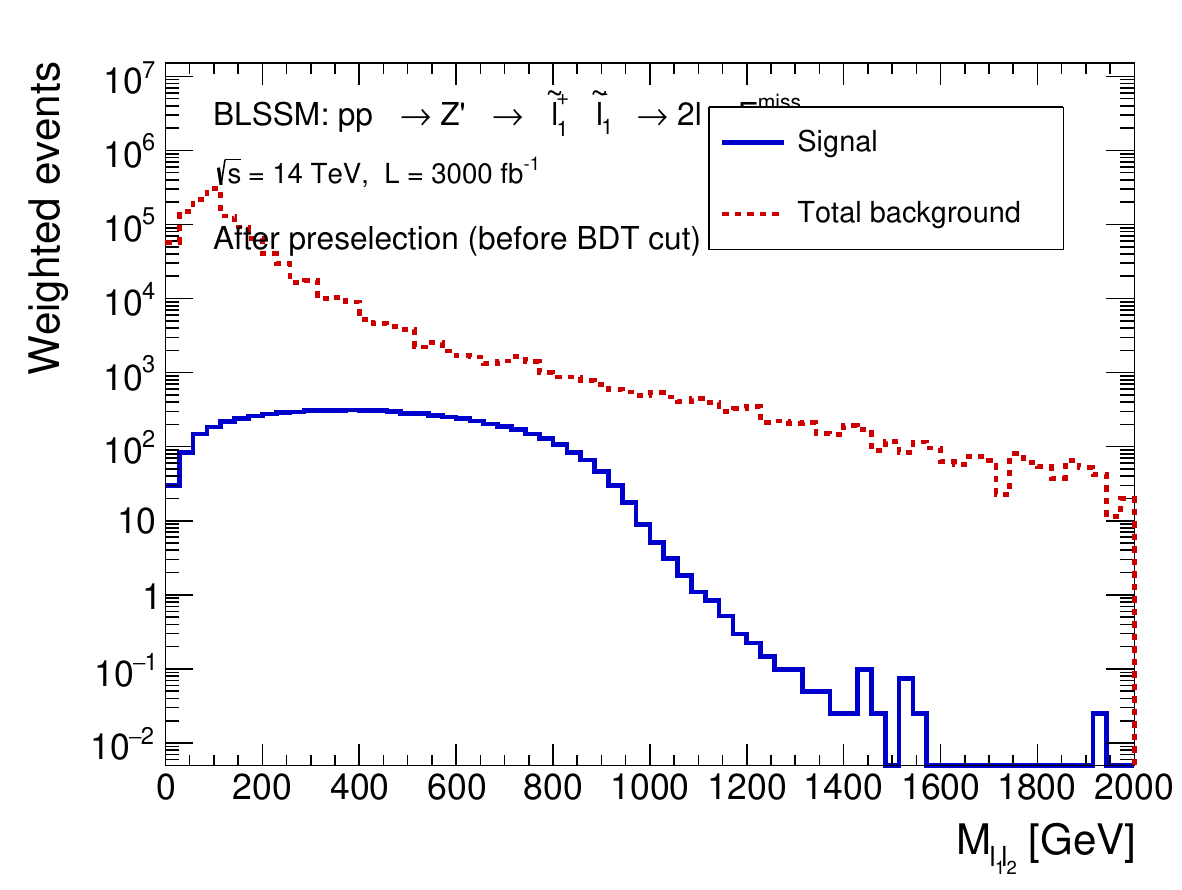}
		\caption{After preselection.}
		\label{fig:mll_preselection_2lMET}
	\end{subfigure}
	\hfill
	\begin{subfigure}{0.32\textwidth}
		\centering
		\includegraphics[width=\linewidth]{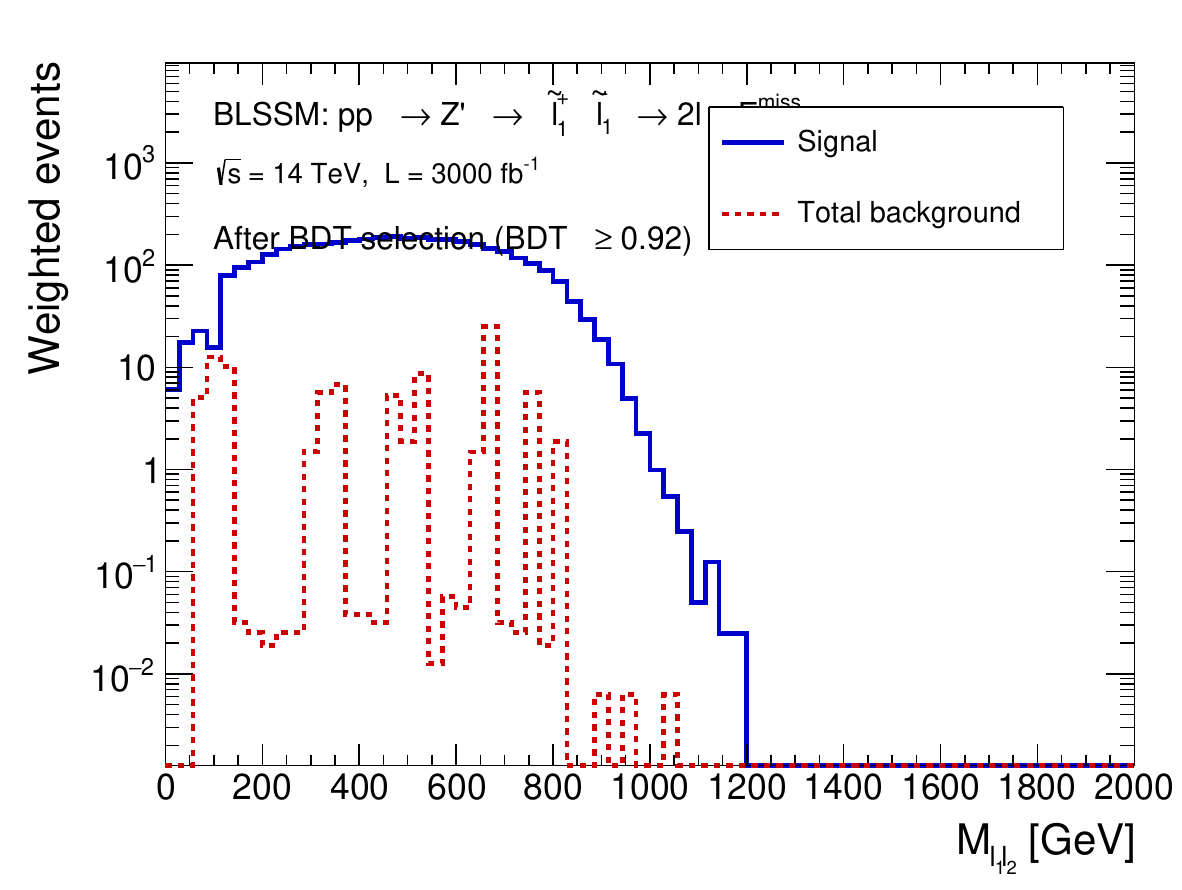}
		\caption{After the optimized BDT selection.}
		\label{fig:mll_BDT_2lMET}
	\end{subfigure}
	
	\caption{Weighted dilepton invariant-mass distributions, $m_{\ell\ell}$,
		for the signal and the SM backgrounds in the
		$2\ell+E_T^{\rm miss}$ final state: before preselection, after the basic
		preselection, and after the optimized BDT selection.}
	\label{fig:mll_evolution_2lMET}
\end{figure}

\begin{figure}[htbp]
	\centering
	
	\begin{subfigure}{0.32\textwidth}
		\centering
		\includegraphics[width=\linewidth]{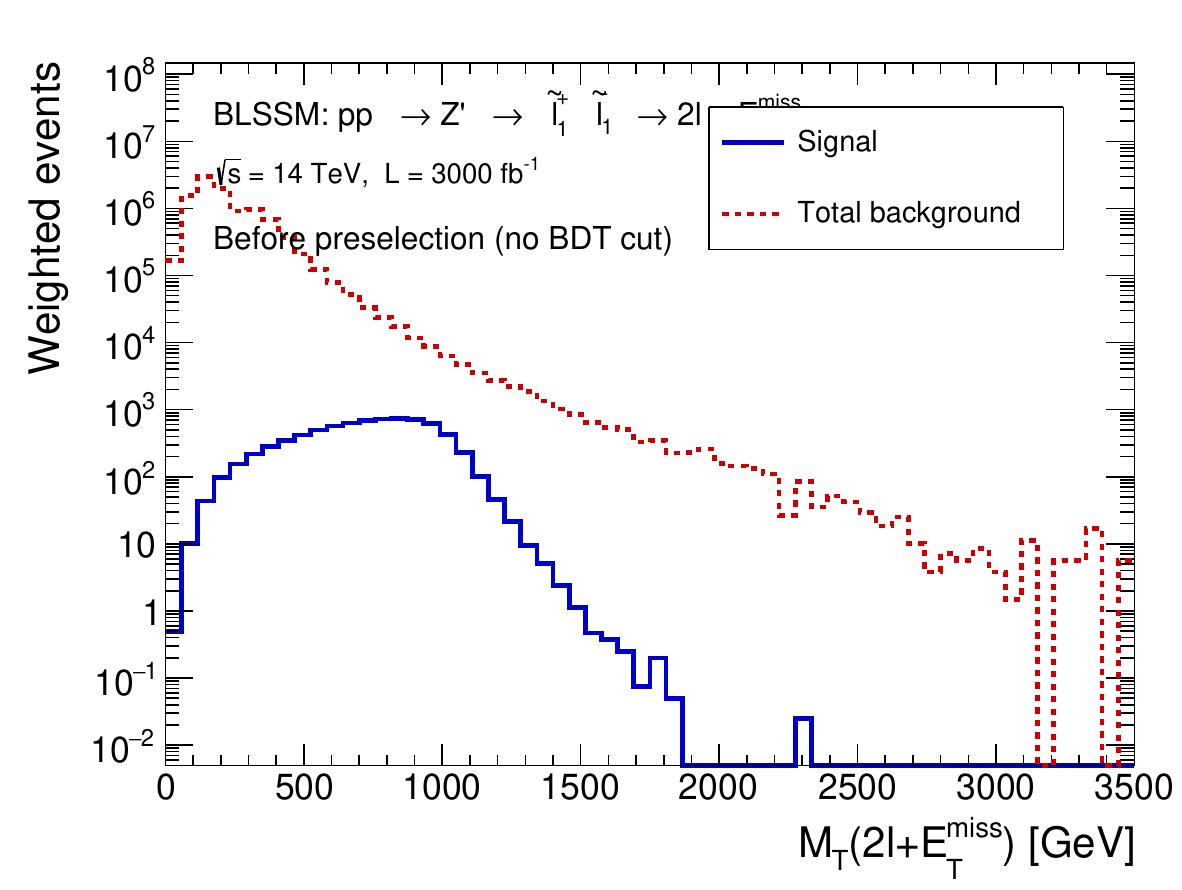}
		\caption{Before preselection.}
		\label{fig:MT_before_2lMET}
	\end{subfigure}
	\hfill
	\begin{subfigure}{0.32\textwidth}
		\centering
		\includegraphics[width=\linewidth]{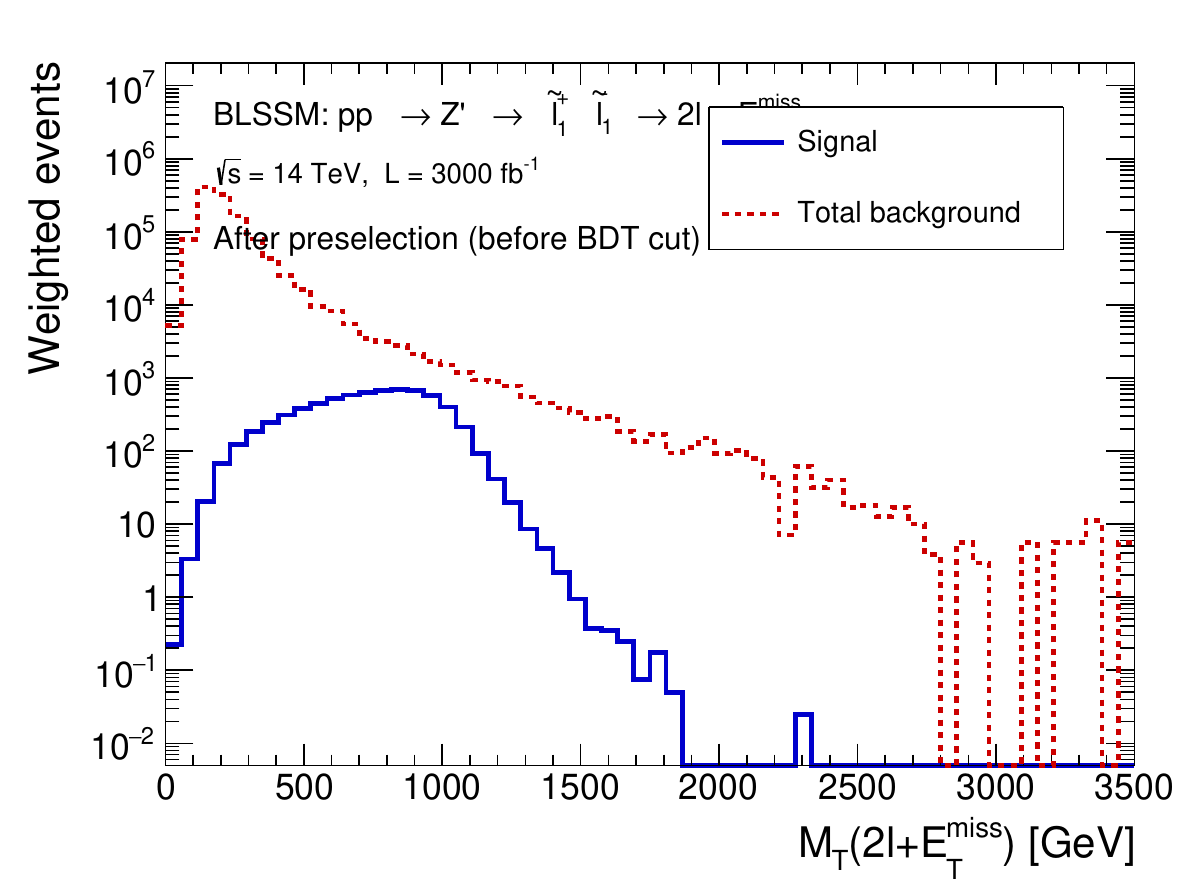}
		\caption{After preselection.}
		\label{fig:MT_preselection_2lMET}
	\end{subfigure}
	\hfill
	\begin{subfigure}{0.32\textwidth}
		\centering
		\includegraphics[width=\linewidth]{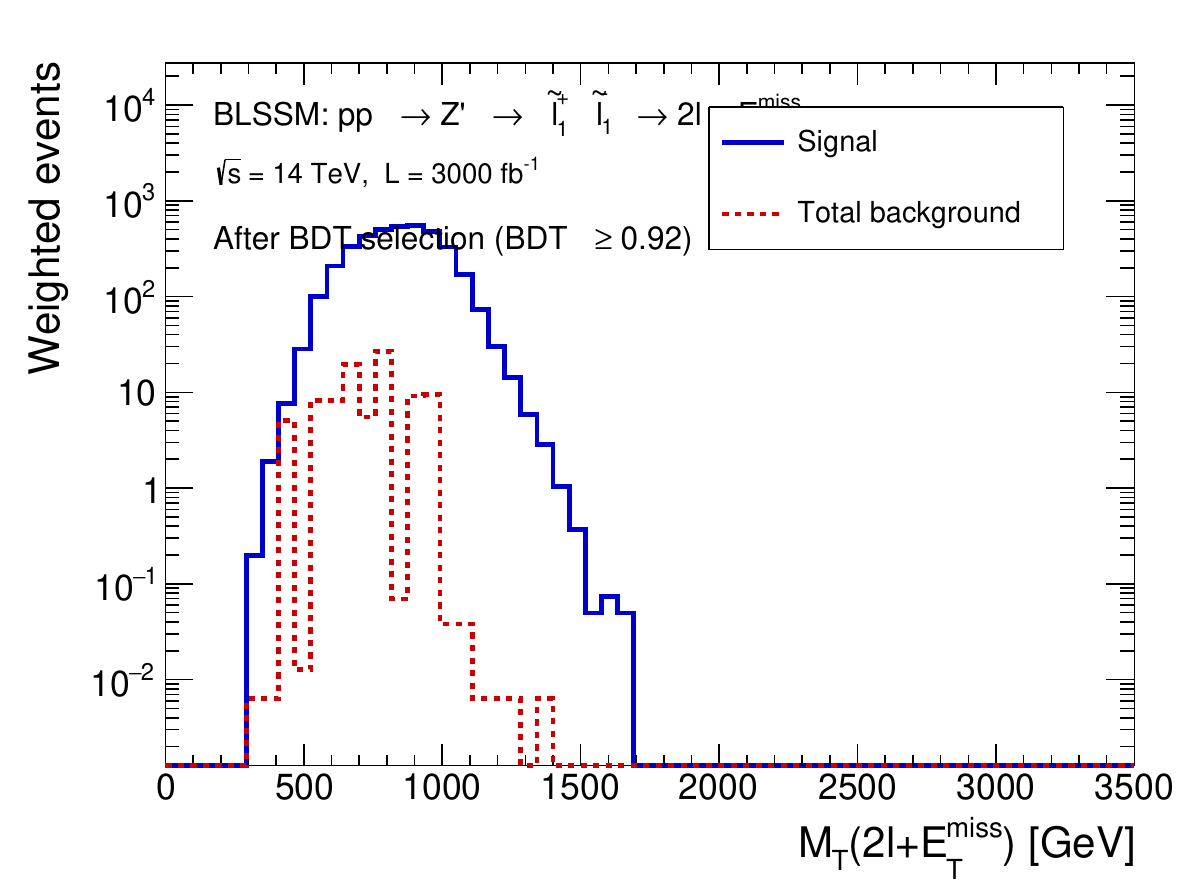}
		\caption{After the optimized BDT selection.}
		\label{fig:MT_BDT_2lMET}
	\end{subfigure}
	
	\caption{Weighted distributions of
		$M_T(2\ell+E_T^{\rm miss})$, used as a proxy observable for the heavy
		$Z'$ resonance, before preselection, after the basic preselection, and
		after the optimized BDT selection.}
	\label{fig:MT_evolution_2lMET}
\end{figure}

\begin{figure}[htbp]
	\centering
	
	\begin{subfigure}{0.32\textwidth}
		\centering
		\includegraphics[width=\linewidth]{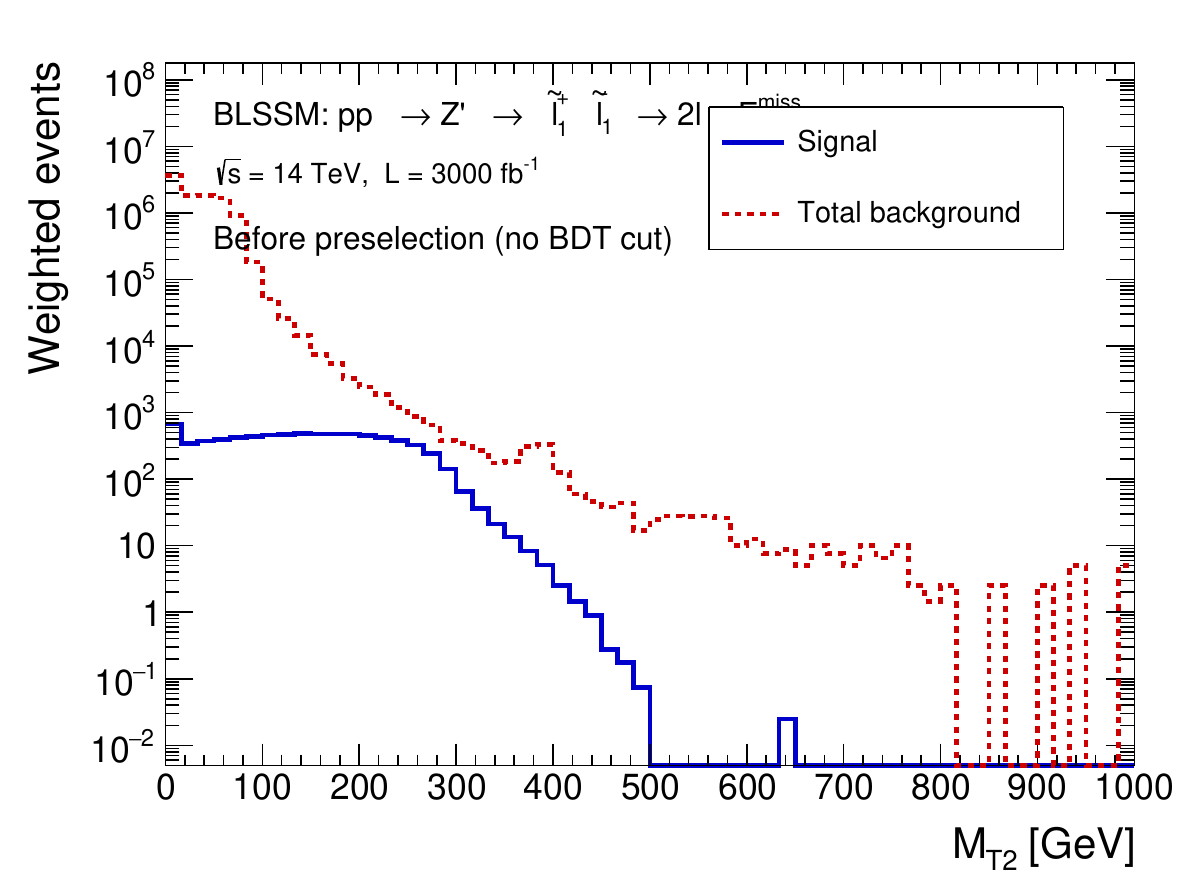}
		\caption{Before preselection.}
		\label{fig:MT2_before_2lMET}
	\end{subfigure}
	\hfill
	\begin{subfigure}{0.32\textwidth}
		\centering
		\includegraphics[width=\linewidth]{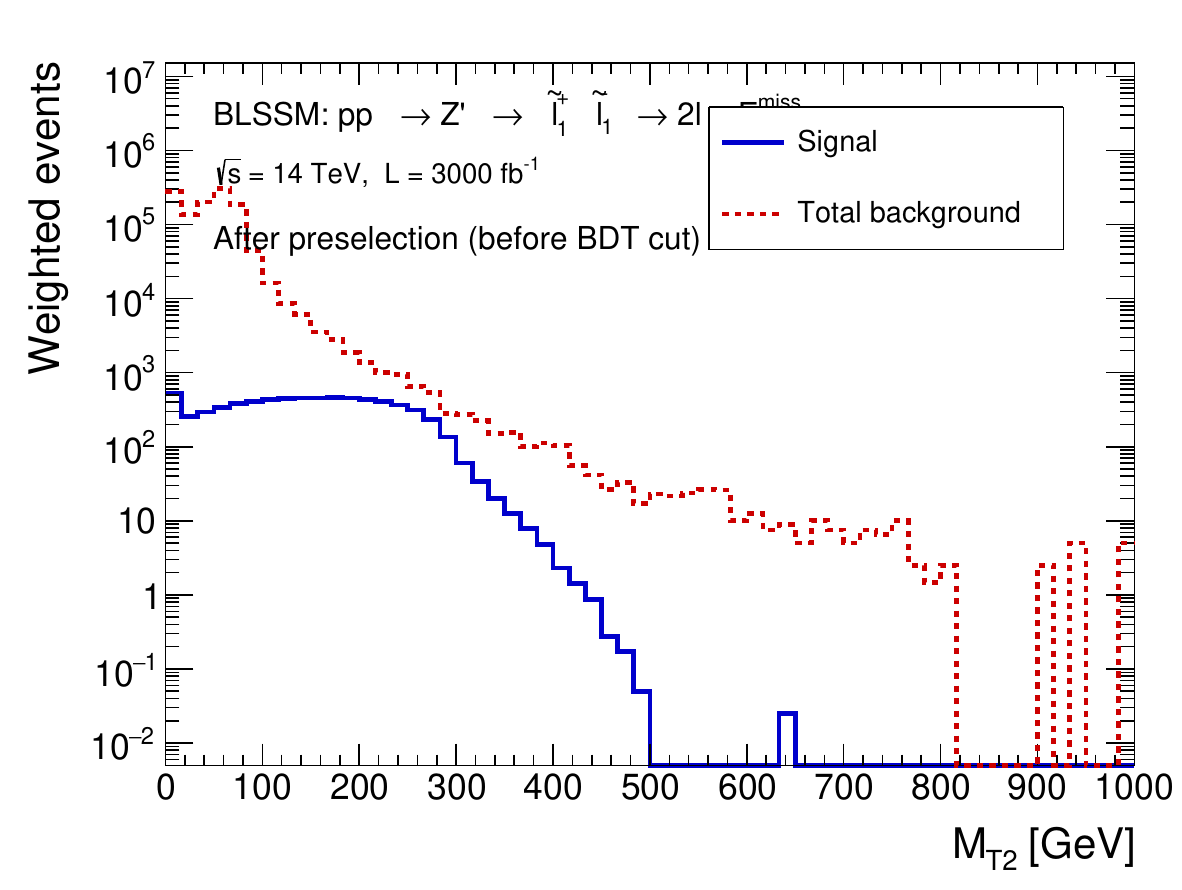}
		\caption{After preselection.}
		\label{fig:MT2_preselection_2lMET}
	\end{subfigure}
	\hfill
	\begin{subfigure}{0.32\textwidth}
		\centering
		\includegraphics[width=\linewidth]{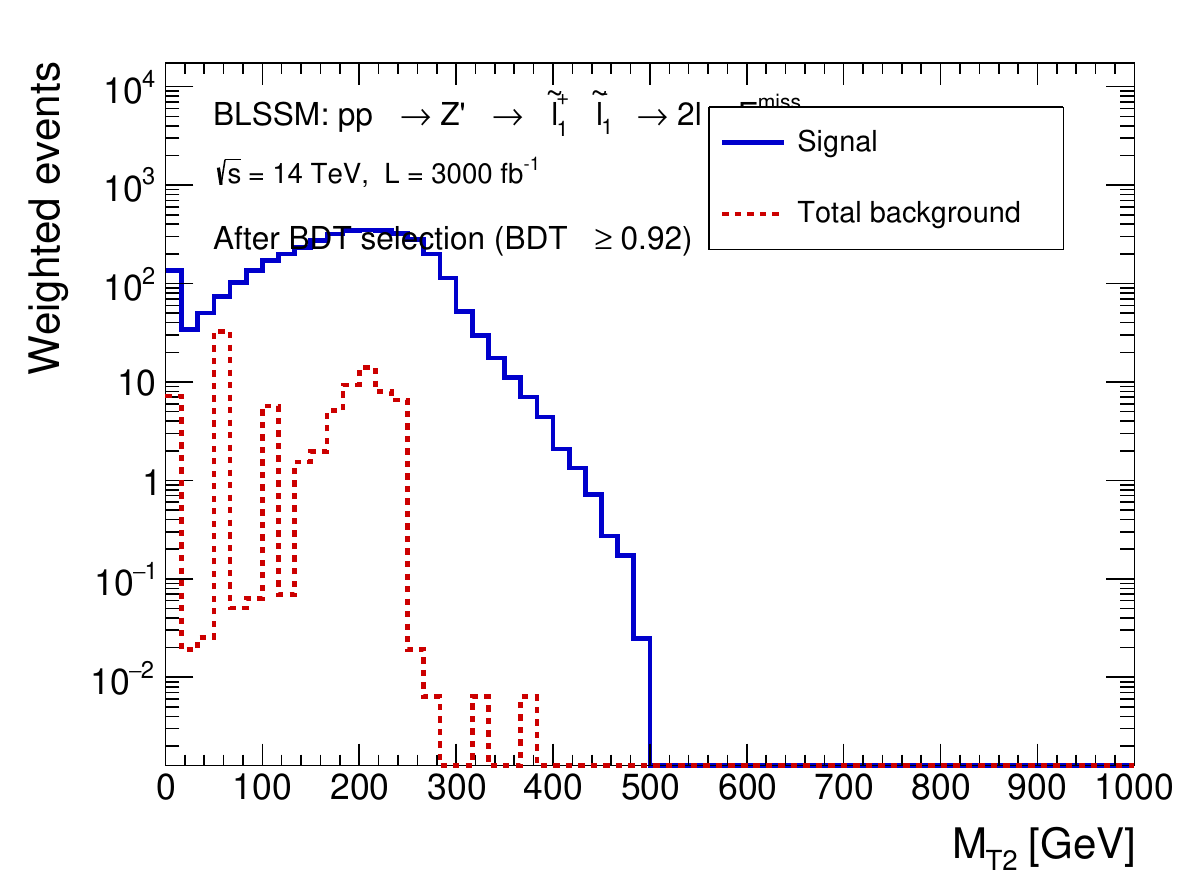}
		\caption{After the optimized BDT selection.}
		\label{fig:MT2_BDT_2lMET}
	\end{subfigure}
	
	\caption{Weighted $M_{T2}$ distributions for the charged-slepton signal
		and the SM backgrounds before preselection, after the basic
		preselection, and after the optimized BDT selection.}
	\label{fig:MT2_evolution_2lMET}
\end{figure}

The evolution of the principal kinematic observables throughout the
analysis is illustrated in
Figs.~\ref{fig:mll_evolution_2lMET}--
\ref{fig:MT2_evolution_2lMET},
where the distributions are shown before the basic preselection,
after the preselection requirements,
and after the optimized BDT selection.

Figure~\ref{fig:mll_evolution_2lMET}
shows the evolution of the dilepton invariant-mass distribution.
The signal exhibits a significantly harder
$m_{\ell\ell}$ spectrum than the SM backgrounds,
a feature that becomes increasingly pronounced after the optimized BDT
selection.
%


The transverse mass
$M_T(2\ell+E_T^{\rm miss})$,
shown in
Fig.~\ref{fig:MT_evolution_2lMET},
provides an effective proxy for the heavy $Z'$ resonance.
The signal extends toward considerably larger transverse masses than
the SM backgrounds, leading to excellent discrimination in
the high-mass region.


The corresponding
$M_{T2}$
distributions are presented in
Fig.~\ref{fig:MT2_evolution_2lMET}.
Since this observable is specifically designed for pair-produced
particles decaying into visible and invisible final states,
it proves particularly effective in separating the slepton signal from
the dominant backgrounds.

The trained BDT exhibits excellent discrimination between the signal
and the SM backgrounds,
yielding an area under the ROC curve{\color{blue}:} 
$ \mathrm{AUC}=0.979,$
while the KS test yields probabilities of $89.3\%$ and
$22.7\%$ for the signal and background samples, respectively,
indicating no significant evidence of overtraining.

The expected sensitivity is evaluated using the Asimov significance $Z_A$
defined in Eqs.~\eqref{eq:Asimov_withSys}, considering background systematic
uncertainties of $\delta_B=0\%,\,5\%,\,10\%,$ and $20\%$.

{The expected $Z_A$ remains above the conventional
$5\sigma$ discovery threshold for all the considered background systematic
uncertainties, including the largest value of
$\delta_B=20\%$, with $Z_A$ increasing when $\delta_B$ decreases, demonstrating the robust
discovery potential of the $2\ell+E_T^{\rm miss}$ channel for the
benchmark scenario considered.}

These results demonstrate that the
$2\ell+E_T^{\rm miss}$
final state constitutes a powerful probe of resonant charged-slepton
production through a heavy $Z'$ boson.
The combination of optimized event selection and multivariate analysis
efficiently suppresses the SM backgrounds while preserving
a large fraction of the signal events. Although the expected significance
is reduced when systematic uncertainties on the background prediction
are included, it remains sufficiently high to provide strong discovery
potential over the range of systematic uncertainties considered.
This channel therefore represents the primary discovery mode for the
charged-slepton sector of the BLSSM at the HL-LHC.

\subsection{Slepton {Cascade(s)} Decays:
$\mathbf{{4(6)}\ell+E_T^{\rm miss}}$}
\label{subsec:slepton4l}

In addition to the direct slepton decay discussed in the previous
subsection, { one (two)} resonantly produced charged sleptons may undergo cascade
decays through heavier electroweakinos, giving rise to multilepton final
states. Of particular interest is the
\begin{equation}
{4(6)}\ell+E_T^{\rm miss}
\end{equation}
signature, characterized by {four (six)} isolated charged leptons and sizable
missing transverse momentum carried by the invisible lightest
neutralinos. At the parton level, the production process is initiated by
\begin{equation}
pp\rightarrow Z'
\rightarrow\tilde{\ell}_1^+\tilde{\ell}_1^-,
\label{eq:slepton_cascade_prod}
\end{equation}
followed by cascade decays of { one (two)} slepton(s) through heavier
electroweakino states. The additional decay steps generate the
multilepton final state but, at the same time, introduce substantial
branching-ratio suppression relative to the direct
$2\ell+E_T^{\rm miss}$ mode.

The {four(six)}-lepton final state has the important experimental advantage of
a relatively small SM background. The dominant irreducible
backgrounds, possibly to mix with our signal, arise mainly from multiboson and top-associated processes,
\begin{equation}
ZZ,\qquad
WZZ,\qquad
ZZZ,\qquad
t\bar t Z,
\end{equation}
while reducible contributions from $t\bar t$, $WZ$, and $Z+$jets can be
strongly suppressed by requiring {four(six)} isolated charged leptons, sizable
missing transverse momentum, and a veto on tagged $b$ jets. Thus, from
the point of view of background rejection, this topology is considerably
cleaner than the direct $2\ell+E_T^{\rm miss}$ slepton channel.

The advantage of the clean final state is, however, offset by the small
signal rate. In contrast to the direct decay
$\tilde{\ell}_1^\pm\to\ell^\pm\tilde{\chi}_1^0$, the multilepton
signature requires additional cascade decays through heavier
electroweakinos. Actually one can anticipate that the $6\ell$-process would be negligible compared to the $4\ell$-process since it implies an additional cascade, and we checked that for our benchmarks the former (latter) cross section gives $\sigma \sim 3 \times 10^{-11} (10^{-5}) fb^{-1}$ which represents a fall by $5$ orders of magnitude,  justifying dropping henceforth the two-cascades process. Symbolically,  the observable cross section scales as
\begin{align}
\sigma_{{4}\ell+E_T^{\rm miss}}
\sim\;&
\sigma(pp\to Z')
\,{\rm BR}
   (Z'\to\tilde{\ell}_1^+\tilde{\ell}_1^-)
\\
& \times {\rm BR}(\tilde{\ell}_1\to{{\ell}_1+\rm E_T^{\rm miss}}) {\rm BR}(\tilde{\ell}_1\to{{\ell}_1+\rm cascade}) 
{\rm BR}({\rm cascade}\to2\ell+E_T^{\rm miss}). \nonumber
\label{eq:slepton4lrate}
\end{align}
The successive branching fractions therefore substantially reduce the
number of observable signal events.

Candidate events are selected by requiring {four} isolated charged
leptons, sizable $E_T^{\rm miss}$, and no tagged $b$ jets. A multivariate
analysis analogous to that employed for the direct slepton channel was
also investigated, using observables characterizing the individual
leptons, the {four}-lepton system, the missing transverse momentum, and
the global event topology. Despite the efficient suppression of the SM
background, the small signal cross section remains the limiting factor.

For the benchmark scenario considered here, even after optimizing the
event selection and multivariate discriminator, the maximum statistical
significance at the HL-LHC remains below unity,
\begin{equation}
{\cal Z}_{{4}\ell+E_T^{\rm miss}} < 1,
\qquad
{\cal L}=3000~{\rm fb}^{-1}.
\label{eq:slepton4lsignificance}
\end{equation}
The cascade ${4}\ell+E_T^{\rm miss}$ signature therefore cannot provide a
statistically meaningful probe of $Z'$-mediated charged-slepton
production at the HL-LHC for the benchmark considered here. This is a
rate limitation rather than a background limitation: the final state is
experimentally very clean, but too few signal events survive the
successive cascade branching fractions.

It is instructive to compare this result with the multilepton signal
from $Z'$-mediated production of heavy right-handed neutrinos studied
in Ref.~\cite{Ezzat:2024twd}. In that case, the
process
\begin{equation}
pp\to Z'\to N_R N_R
\end{equation}
followed by the leptonic decays of the heavy neutrinos gives rise to
multilepton final states with missing transverse momentum and was found
to have substantially better discovery prospects. The difference is
primarily associated with the signal rate: the slepton cascade studied
here suffers from additional branching-ratio suppressions through the
intermediate supersymmetric states, despite sharing the advantage of a
clean multilepton final state.

This comparison illustrates that a clean experimental signature does
not necessarily imply a competitive discovery channel. In the charged
slepton sector considered here, the cascade
${4}\ell+E_T^{\rm miss}$ mode is limited by its small production rate,
whereas the direct $2\ell+E_T^{\rm miss}$ channel retains a much larger
signal yield and therefore constitutes the primary HL-LHC probe of
$Z'$-mediated slepton production.


\subsection{Right-Handed Sneutrino Channels}
\label{subsec:sneutrino}

We now turn to the complementary production of right-handed sneutrinos
through the resonant decay of the $Z'$ boson,
\begin{equation}
pp\rightarrow Z'
\rightarrow
\tilde{\nu}_R \tilde{\nu}_I.
\end{equation}
For the benchmark scenario considered here, the right-handed sneutrinos
decay predominantly through charged leptons and charginos,
\begin{equation}
\tilde{\nu}_{R{, I}}
\rightarrow
\ell^\mp\tilde{\chi}_1^\pm,
\qquad
\tilde{\chi}_1^\pm
\rightarrow
W^\pm\tilde{\chi}_1^0,
\end{equation}
leading to
\begin{equation}
pp\rightarrow Z'
\rightarrow
\tilde{\nu}_R \tilde{\nu}_I
\rightarrow
\ell^+\ell^-W^+W^-+E_T^{\rm miss}.
\label{eq:sneutrino_master}
\end{equation}
Depending on the subsequent decays of the two $W$ bosons, this process
gives rise to three characteristic final states,
\begin{align}
W^+W^- \to \ell^+\nu\,\ell^-\bar\nu:
&\qquad
4\ell+E_T^{\rm miss},
\nonumber\\
W^+W^- \to \ell\nu\,jj:
&\qquad
3\ell+2j+E_T^{\rm miss},
\nonumber\\
W^+W^- \to jj\,jj:
&\qquad
2\ell+4j+E_T^{\rm miss}.
\label{eq:sneutrino_finalstates}
\end{align}

These three channels exhibit a characteristic trade-off between
experimental cleanliness and signal rate. The fully leptonic
$4\ell+E_T^{\rm miss}$ channel has the smallest SM background, but its
rate is strongly suppressed by the leptonic branching fractions of both
$W$ bosons. The semileptonic
$3\ell+2j+E_T^{\rm miss}$ channel benefits from a larger branching
fraction, but its sensitivity remains limited for the benchmark
considered here. In both cases, the optimized statistical significance
is found to be too small to provide a competitive HL-LHC probe.
We therefore do not pursue detailed multivariate analyses of these two
channels.

In contrast, when both $W$ bosons decay hadronically,
\begin{equation}
W^\pm\rightarrow jj,
\end{equation}
the substantially larger hadronic branching fraction leads to a
sufficiently enhanced signal rate. The resulting
\begin{equation}
2\ell+4j+E_T^{\rm miss}
\end{equation}
signature is therefore the only right-handed sneutrino topology among
those considered here that provides appreciable sensitivity at the
HL-LHC. We consequently focus the remainder of this subsection on this
channel.


The signal process is
\begin{equation}
pp\rightarrow Z'
\rightarrow
\tilde{\nu}_{R}\tilde{\nu}_{I}
\rightarrow
\ell^+\ell^-\tilde{\chi}_1^+\tilde{\chi}_1^-
\rightarrow
\ell^+\ell^-W^+W^-\tilde{\chi}_1^0\tilde{\chi}_1^0 \xrightarrow{W^+W^-\rightarrow 4j} 2\ell+4j+E_T^{\rm miss}.
\end{equation}

{
The dominant SM backgrounds considered in this analysis are the
inclusive $2\ell+4j$ sample, dileptonic $t\bar t+2j$ production,
and $W+$jets events with fake leptons,
\begin{equation}
	\mathrm{SM}~2\ell+4j,\qquad
	t\bar{t}+2j~(\mathrm{dileptonic}),\qquad
	W+\mathrm{jets}~(\mathrm{fake~leptons}).
\end{equation}
	
}

The large $t\bar t$ contribution can be efficiently reduced by a
$b$-jet veto, while the signal topology is further characterized by
four energetic jets, two isolated charged leptons, and sizable
$E_T^{\rm miss}$.

{The preselection requires exactly two isolated charged leptons and at
	least four reconstructed jets, with
	$p_T(\ell_1)>25~{\rm GeV}$ and $p_T(\ell_2)>20~{\rm GeV}$.
	Events are further required to satisfy
	$E_T^{\rm miss}>50~{\rm GeV}$, while a $b$-jet veto is imposed to
	suppress backgrounds from top-quark production.}
	
	{
	The event yields after the preselection and the optimized BDT
	selection are summarized in Table~\ref{tab:cutflow_2l4jMET}.

	\begin{table}[htbp]
		\centering
		\begin{tabular}{lccccc}
			\hline
			Process &
			$\sigma$ (fb) &
			Generated &
			After preselection &
			After BDT &
			Expected events \\
			\hline
			
			Signal
			& 0.173 & 300000 & 21555 & 21175 & 36.63 \\
			
			\hline
			
			SM $2\ell+4j$
			& 0.67
			& 200000
			& 31442
			& 2
			& $< O(1)$ \\
			
			$t\bar t+2j$ (dileptonic)
			& 23.0
			& 300000
			& 6074
			& 66& 15.18\\
			
			$W+$jets (fake leptons)
			& 782.5
			& 68323
			& 2
			& $< O(1)$ & $< O(1)$\\
            \hline
			Total background &&&&& 15.20 \\
			\hline
		\end{tabular}
			\caption{Production cross sections, generated Monte Carlo events,
				surviving events after the basic preselection and the optimized BDT
				selection, together with the corresponding expected event yields at
				the HL-LHC with an integrated luminosity of $3000~\mathrm{fb}^{-1}$.
				For the SM background samples, the last two columns report the combined
				background contribution after the BDT selection.}
			\label{tab:cutflow_2l4jMET}
	\end{table}
}

The invariant masses of the jet pairs and the reconstructed hadronic
$W$ candidates provide additional information on the hadronic decay
topology of the signal.

{The multivariate analysis exploits a broad set of kinematic observables
	characterizing the leptons, jets, missing transverse momentum,
	reconstructed hadronic $W$ candidates, and the global event activity.
	Among the most discriminating variables are $L_T$, $p_T(\ell_1)$,
	$E_T^{\rm miss}$, $\Delta\phi(\ell_2,E_T^{\rm miss})$,
	$m_{W\ell}^{\rm min}$, $p_T(\ell_2)$, $m_{\ell\ell}$,
	and $m_{2\ell4j}$. The ten highest-ranked BDT input variables are
	summarized in Table~\ref{tab:ranking_2l4j}.}


\begin{table}[htbp]
	\centering
	\begin{tabular}{ccc}
		\hline
		Rank & Variable & Importance \\
		\hline
		1  & $L_T$                                   & 0.0838 \\
		2  & $p_T(\ell_1)$                           & 0.0760 \\
		3  & $E_T^{\rm miss}$                        & 0.0645 \\
		4  & $\Delta\phi(\ell_2,E_T^{\rm miss})$     & 0.0538 \\
		5  & $m_{W\ell}^{\rm min}$                   & 0.0496 \\
		6  & $p_T(\ell_2)$                           & 0.0452 \\
		7  & $\Delta\phi_{\ell\ell}$                 & 0.0387 \\
		8  & $m_{\ell\ell}$                          & 0.0306 \\
		9  & $m_{2\ell4j}$                           & 0.0288 \\
		10 & $p_T(j_1)$                              & 0.0280 \\
		\hline
	\end{tabular}
		\caption{Ranking of the most important BDT input variables for the
		$2\ell+4j+E_T^{\rm miss}$ analysis.}
	\label{tab:ranking_2l4j}
\end{table}

\begin{figure}[htbp]
	\centering
	
	\begin{subfigure}{0.48\textwidth}
		\centering
		\includegraphics[width=\linewidth]{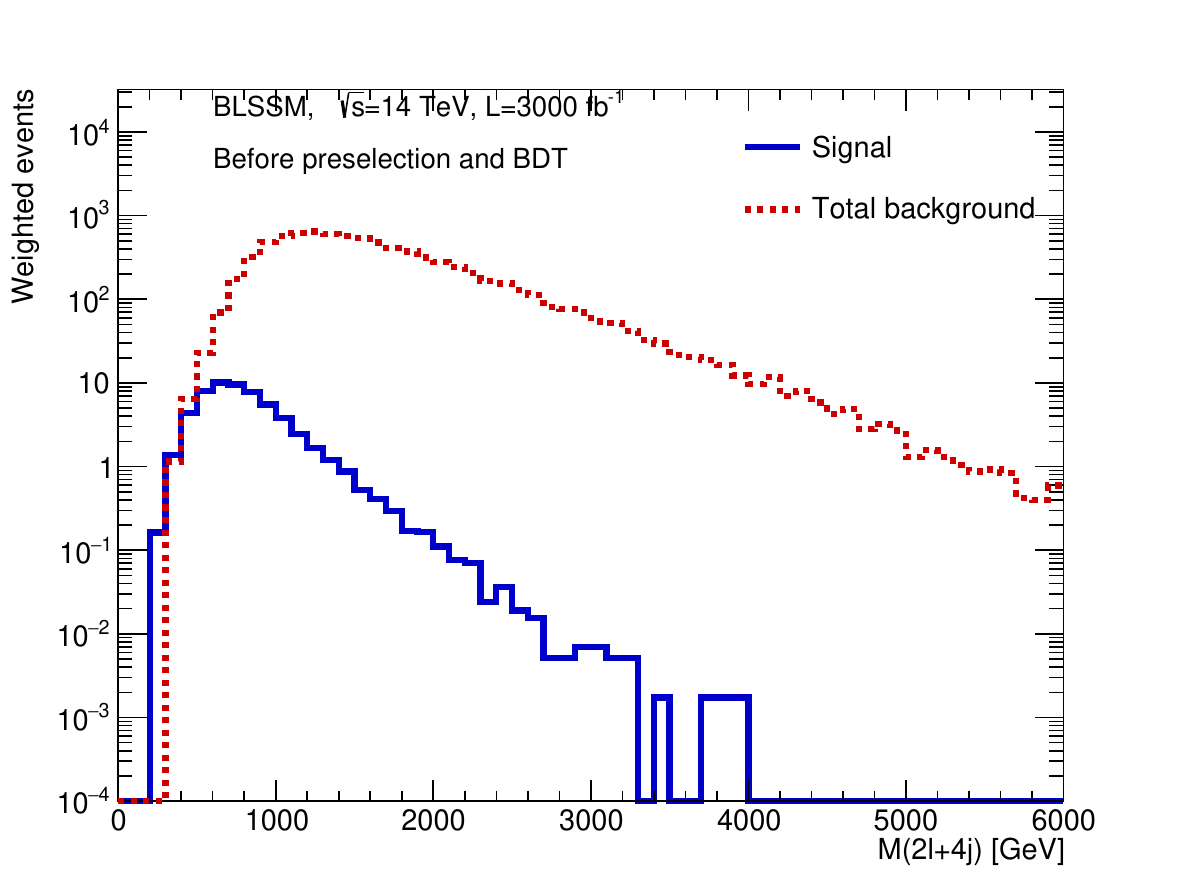}
	\end{subfigure}
	\hfill
	\begin{subfigure}{0.48\textwidth}
		\centering
		\includegraphics[width=\linewidth]{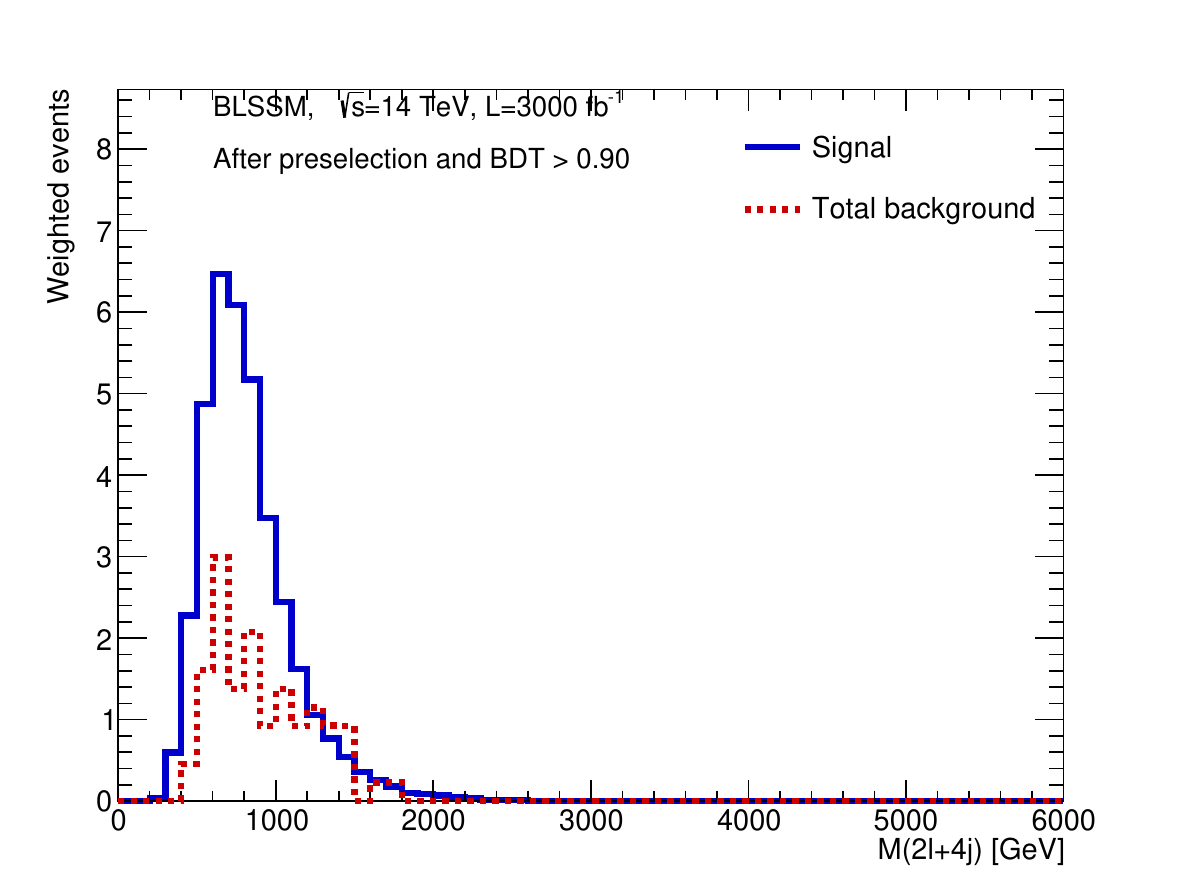}
	\end{subfigure}
	
	\caption{{Invariant-mass distribution of the visible $2\ell+4j$ system,
		$M_{2\ell4j}$, for the signal and the SM backgrounds before the event
		selection (left) and after applying the preselection and the optimized
		BDT requirement (right).}}
	\label{fig:2l4j_2l4jMET}
\end{figure}
\begin{figure}[htbp]
	\centering
	
	\begin{subfigure}{0.48\textwidth}
		\centering
		\includegraphics[width=\linewidth]{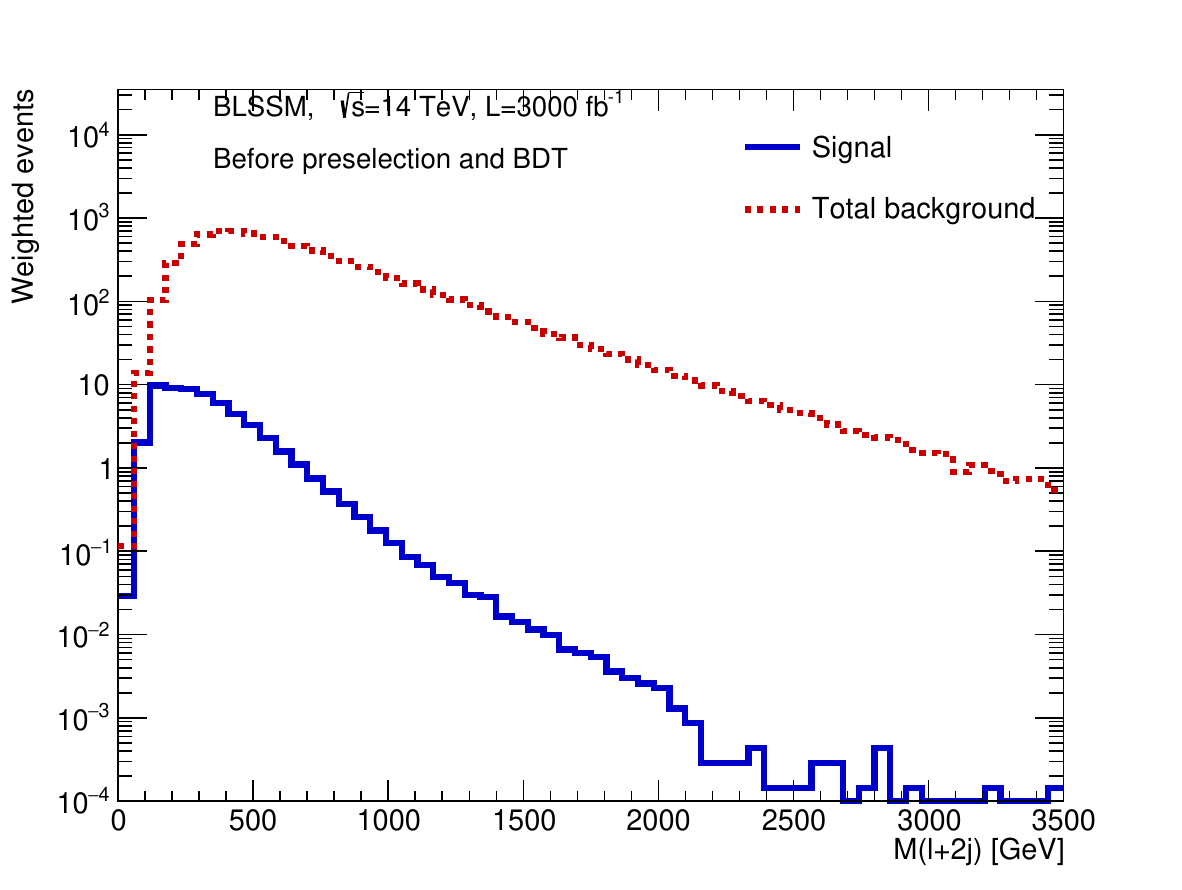}
	\end{subfigure}
	\hfill
	\begin{subfigure}{0.48\textwidth}
		\centering
		\includegraphics[width=\linewidth]{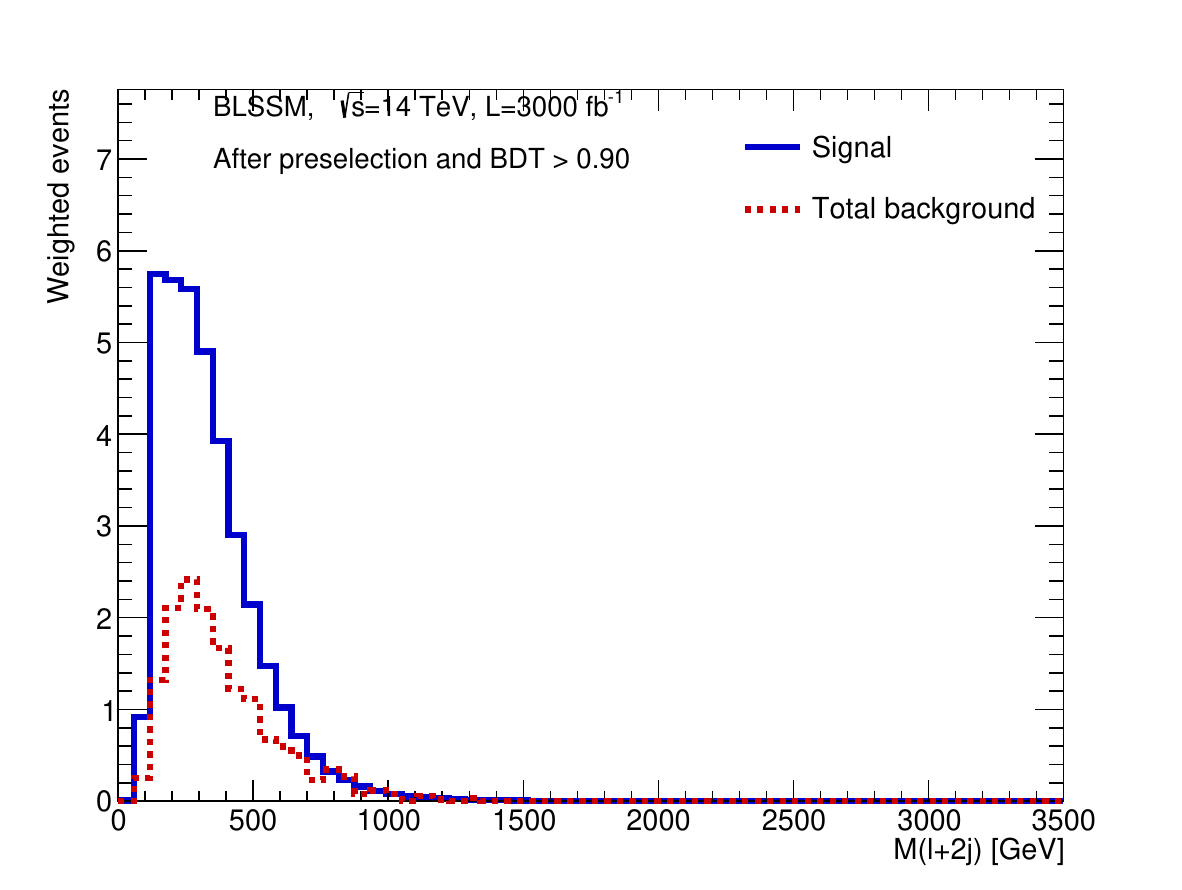}
	\end{subfigure}
	
\caption{{Invariant-mass distribution of the $\ell+2j$ one-branch system,
	$M_{\ell2j}$, for the signal and the SM backgrounds before the event
	selection (left) and after applying the preselection and the optimized
	BDT requirement (right). All possible lepton--dijet combinations are
	included in the reconstruction.}}
\label{fig:1l2j_2l4jMET}
\end{figure}

{ Figures~\ref{fig:2l4j_2l4jMET} and
	\ref{fig:1l2j_2l4jMET} illustrate the effect of the optimized event
	selection on the invariant-mass distributions. Before the selection,
	the signal is largely obscured by the SM background, whereas the
	application of the preselection and BDT requirement strongly suppresses
	the background while retaining a sizeable signal contribution. The
	$M_{2\ell4j}$ distribution probes the visible mass scale of the complete
	decay topology, while the one-branch $M_{\ell2j}$ combinations provide complementary
	information on the intermediate sneutrino decay kinematics.} As to KS test, and like the slepton case, it yields 
probabilities of $58\% (21.6\%)$ for the signal (background),
indicating no significant evidence of overtraining.

\begin{figure}[h!]
	\centering
	\includegraphics[width=0.6\textwidth]
	{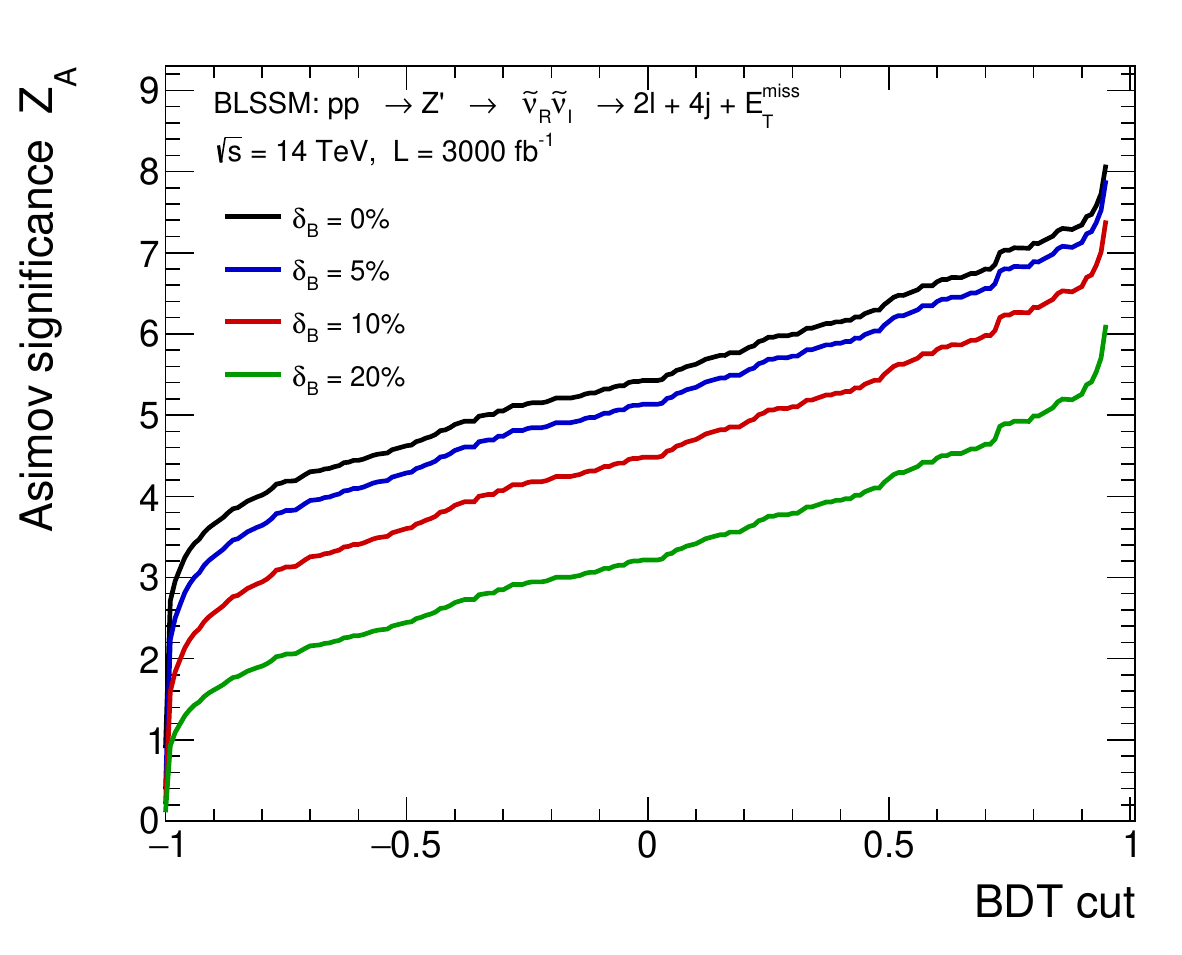}
	\caption{Asimov significance $Z_A$ as a function of the BDT selection
		threshold for the $2\ell+4j+E_T^{\rm miss}$ final state at
		$\sqrt{s}=14~\mathrm{TeV}$ and an integrated luminosity of
		$3000~\mathrm{fb}^{-1}$. The curves correspond to background
		systematic uncertainties of $\delta_B=0\%$, $5\%$, $10\%$, and
		$20\%$.}
	\label{fig:significance_2l4jMET}
\end{figure}

{

Figure~\ref{fig:significance_2l4jMET} shows the $Z_A$ significance as
a function of the BDT selection threshold for several representative
values of the systematic uncertainty on the total background yield.
}%
The optimal BDT selection is determined separately for each assumed
systematic uncertainty. As expected, the significance decreases as the
background systematic uncertainty increases; nevertheless, the BDT
selection retains substantial sensitivity, above the $5\sigma$ level, for all the
uncertainties considered.

Actually, after the optimized event selection and BDT requirement, the
$2\ell+4j+E_T^{\rm miss}$ channel retains sufficient signal statistics
to achieve an observable significance at the HL-LHC. This is in marked
contrast to the fully leptonic and semileptonic sneutrino modes, whose
smaller branching fractions result in statistically suppressed signals.

The comparison among the three channels therefore demonstrates that,
for the benchmark considered here, the larger hadronic branching
fraction of the $W$ bosons more than compensates for the increased SM
background. Consequently, the
$2\ell+4j+E_T^{\rm miss}$ final state constitutes the most promising
HL-LHC probe of $Z'$-mediated right-handed sneutrino production.

\newpage
\section{Conclusions}
\label{sec:conclusion}

In this work, we have investigated $Z'$-mediated production of charged
sleptons and right-handed sneutrinos in the BLSSM at the HL-LHC.
The additional $U(1)_{B-L}$ interaction provides a resonant production
mechanism that can significantly enhance the sensitivity to these
supersymmetric scalar states.

For charged sleptons, the direct decay channel
\begin{equation}
pp\to Z'\to\tilde{\ell}^{+}\tilde{\ell}^{-}
\to 2\ell+E_T^{\rm miss}
\end{equation}
provides the strongest sensitivity. Despite sizeable SM backgrounds,
the resonant enhancement of the slepton production rate, combined with
optimized event selection and a BDT analysis, leads to an expected
statistical significance well above the $5\sigma$ discovery threshold
for the benchmark considered at an integrated luminosity of
$3000~{\rm fb}^{-1}$. In contrast, the cleaner
${4}\ell+E_T^{\rm miss}$ cascade channel suffers from substantial
branching-ratio suppression, resulting in a significance below unity
and therefore limited sensitivity at the HL-LHC.

Right-handed sneutrino production,
$pp\to Z'\to\tilde{\nu}_R\tilde{\nu}_R^{*}$, leads to distinct
multilepton and multijet signatures through cascade decays involving
charginos and neutralinos. Among the channels considered, the fully
leptonic and semileptonic modes have suppressed sensitivity, whereas
the $2\ell+4j+E_T^{\rm miss}$ final state benefits from the larger
hadronic $W$ branching fraction and provides the most promising probe
of the right-handed sneutrino sector.

Our results therefore identify
$2\ell+E_T^{\rm miss}$ and $2\ell+4j+E_T^{\rm miss}$ as the leading
HL-LHC signatures of $Z'$-mediated charged-slepton and right-handed
sneutrino production, respectively. Together with previous searches for
$Z'$-mediated right-handed neutrinos, these channels provide
complementary probes of the fermionic and scalar sectors associated
with the $B-L$ symmetry.

\section*{Acknowledgements}
N. C. acknowledges support of the Alexander von Humboldt
Foundation and is grateful for the kind hospitality of the
Bethe Center for Theoretical Physics at the University
of Bonn. K. E. would like to thank M. Ashry for the fruitful discussions.

\bibliographystyle{utphys}
\bibliography{ref}

@article{KHALIL2008374,
  author  = {Khalil, S. and Masiero, A.},
  title   = {Radiative B-L symmetry breaking in supersymmetric models},
  journal = {Physics Letters B},
  volume  = {665},
  number  = {5},
  pages   = {374--377},
  year    = {2008},
  issn    = {0370-2693},
  doi     = {10.1016/j.physletb.2008.06.063},
  url     = {https://doi.org/10.1016/j.physletb.2008.06.063}
}

@article{OLeary:2011yq,
  author        = {O'Leary, Ben and Porod, Werner and Staub, Florian},
  title         = {Mass spectrum of the minimal SUSY B-L model},
  journal       = {JHEP},
  volume        = {05},
  pages         = {042},
  year          = {2012},
  eprint        = {1112.4600},
  archivePrefix = {arXiv},
  primaryClass  = {hep-ph},
  doi           = {10.1007/JHEP05(2012)042}
}

@article{Basso:2011na,
  author        = {Basso, Lorenzo and Belyaev, Alexander and Moretti, Stefano and Shepherd-Themistocleous, Claire},
  title         = {Phenomenology of the minimal B-L extension of the Standard Model at the LHC},
  journal       = {Phys. Rev. D},
  volume        = {83},
  pages         = {055012},
  year          = {2011},
  eprint        = {1106.4462},
  archivePrefix = {arXiv},
  primaryClass  = {hep-ph},
  doi           = {10.1103/PhysRevD.83.055012}
}

@article{Basso:2015pka,
  author        = {Basso, Lorenzo},
  title         = {The Higgs sector of the minimal SUSY B-L model},
  journal       = {Advances in High Energy Physics},
  volume        = {2015},
  pages         = {980687},
  year          = {2015},
  eprint        = {1504.05328},
  archivePrefix = {arXiv},
  primaryClass  = {hep-ph},
  doi           = {10.1155/2015/980687}
}

@article{Ezzat:2024twd,
    author = "Ezzat, Kareem and Chamoun, Nidal and Khalil, Shaaban and Sengupta, Rhitaja",
    title = "{Exploring Z' and right-handed neutrinos in the B-L symmetric SM at the Large Hadron Collider}",
    eprint = "2412.19269",
    archivePrefix = "arXiv",
    primaryClass = "hep-ph",
    doi = "10.1103/x6dp-4tzs",
    journal = "Phys. Rev. D",
    volume = "113",
    number = "1",
    pages = "015009",
    year = "2026"
}

@article{Basso:2009hf,
  author    = {L. Basso and S. Moretti and G. M. Pruna},
  title     = {Phenomenology of the minimal B-L extension of the Standard Model: The Higgs sector},
  journal   = {JHEP},
  volume    = {10},
  year      = {2009},
  pages     = {006},
  eprint    = {0903.4777},
  archivePrefix = {arXiv},
  primaryClass = {hep-ph}
}

@article{Chun:2017spz,
  author    = {E. J. Chun and S. K. Kang and D. Y. Kim and J. Park},
  title     = {Search for $Z'$ in Type-I Seesaw Models},
  journal   = {JHEP},
  volume    = {11},
  year      = {2017},
  pages     = {036},
  eprint    = {1708.00023},
  archivePrefix = {arXiv},
  primaryClass = {hep-ph}
}

@article{DelleRose:2017ukx,
    author = "Delle Rose, Luigi and Khalil, Shaaban and King, Simon J. D. and Kulkarni, Suchita and Marzo, Carlo and Moretti, Stefano and Un, Cem S.",
    title = "{Sneutrino Dark Matter in the BLSSM}",
    eprint = "1712.05232",
    archivePrefix = "arXiv",
    primaryClass = "hep-ph",
    doi = "10.1007/JHEP07(2018)100",
    journal = "JHEP",
    volume = "07",
    pages = "100",
    year = "2018"
}

@article{Huitu:2008gf,
    author = "Huitu, Katri and Khalil, Shaaban and Okada, Hiroshi and Rai, Santosh Kumar",
    title = "{Signatures for right-handed neutrinos at the Large Hadron Collider}",
    eprint = "0803.2799",
    archivePrefix = "arXiv",
    primaryClass = "hep-ph",
    doi = "10.1103/PhysRevLett.101.181802",
    journal = "Phys. Rev. Lett.",
    volume = "101",
    pages = "181802",
    year = "2008"
}

@article{Baumgart:2006pa,
    author = "Baumgart, Matthew and Hartman, Thomas and Kilic, Can and Wang, Lian-Tao",
    title = "{Discovery and measurement of sleptons, binos, and winos with a Z'}",
    eprint = "hep-ph/0608172",
    archivePrefix = "arXiv",
    doi = "10.1088/1126-6708/2007/11/084",
    journal = "JHEP",
    volume = "11",
    pages = "084",
    year = "2007"
}

@article{Corcella:2014xwa,
    author = "Corcella, Gennaro",
    title = "{Phenomenology of supersymmetric Z' decays at the Large Hadron Collider}",
    journal = "Eur. Phys. J. C",
    volume = "75",
    number = "6",
    pages = "264",
    year = "2015",
    doi = "10.1140/epjc/s10052-015-3459-9"
}

@article{Abdallah:2015uba,
    author = "Abdallah, W. and Fiaschi, J. and Khalil, S. and Moretti, S.",
    title = "{Z'-induced Invisible Right-handed Sneutrino Decays at the LHC}",
    eprint = "1504.01761",
    archivePrefix = "arXiv",
    primaryClass = "hep-ph",
    doi = "10.1103/PhysRevD.92.055029",
    journal = "Phys. Rev. D",
    volume = "92",
    number = "5",
    pages = "055029",
    year = "2015"
}

@article{Elsayed:2012ec,
    author = "Elsayed, A. and Khalil, S. and Moretti, S. and Moursy, A.",
    title = "{Right-handed sneutrino-antisneutrino oscillations in a TeV scale Supersymmetric B-L model}",
    eprint = "1211.0644",
    archivePrefix = "arXiv",
    primaryClass = "hep-ph",
    doi = "10.1103/PhysRevD.87.053010",
    journal = "Phys. Rev. D",
    volume = "87",
    number = "5",
    pages = "053010",
    year = "2013"
}

@article{Bandyopadhyay:2011es,
    author = "Bandyopadhyay, Priyotosh and Chun, Eung Jin and Park, Jong-Chul",
    title = "{Right-handed sneutrino dark matter in U(1)' seesaw models and its signatures at the LHC}",
    eprint = "1105.1652",
    archivePrefix = "arXiv",
    primaryClass = "hep-ph",
    doi = "10.1007/JHEP06(2011)129",
    journal = "JHEP",
    volume = "06",
    pages = "129",
    year = "2011"
}

@article{Arina:2013zca,
    author = "Arina, Chiara and Cabrera, Maria Eugenia",
    title = "{Multi-lepton signatures at LHC from sneutrino dark matter}",
    doi = "10.1007/JHEP04(2014)100",
    journal = "JHEP",
    volume = "04",
    pages = "100",
    year = "2014"
}

@article{Xia:2018cfz,
    author = "Xia, Li-Gang",
    title = "{Understanding the boosted decision tree methods with the weak-learner approximation}",
    eprint = "1811.04822",
    archivePrefix = "arXiv",
    primaryClass = "physics.data-an",
    month = "11",
    year = "2018"
}

@article{XIA201915,
title = {QBDT, a new boosting decision tree method with systematical uncertainties into training for High Energy Physics},
journal = {Nuclear Instruments and Methods in Physics Research Section A: Accelerators, Spectrometers, Detectors and Associated Equipment},
volume = {930},
pages = {15-26},
year = {2019},
issn = {0168-9002},
doi = {https://doi.org/10.1016/j.nima.2019.03.088},
url = {https://www.sciencedirect.com/science/article/pii/S0168900219304309},
author = {Li-Gang Xia}
}

@article{quinlan1986induction,
  author    = {J. Ross Quinlan},
  title     = {Induction of Decision Trees},
  journal   = {Machine Learning},
  volume    = {1},
  number    = {1},
  pages     = {81--106},
  year      = {1986},
  publisher = {Springer}
}

@article{quinlan1987simplifying,
  author    = {J. Ross Quinlan},
  title     = {Simplifying Decision Trees},
  journal   = {International Journal of Man-Machine Studies},
  volume    = {27},
  number    = {3},
  pages     = {221--234},
  year      = {1987},
  publisher = {Elsevier}
}

@article{hoecker2007tmva,
  author    = {A. H{\"o}cker and P. Speckmayer and J. Stelzer and A. Voss and H. D. Zech},
  title     = {TMVA - Toolkit for Multivariate Data Analysis},
  journal   = {Physics Communications},
  volume    = {168},
  pages     = {107--123},
  year      = {2007},
  doi       = {10.1016/j.physcom.2006.11.005}
}

@article{Alwall:2014hca,
    author = "Alwall, Johan and Frederix, Rikkert and Frixione, Stefano
              and Hirschi, Valentin and Maltoni, Fabio and Mattelaer, Olivier
              and Shao, Hua-Sheng and Stelzer, Tim and Torrielli, Paolo
              and Zaro, Marco",
    title = "{The automated computation of tree-level and next-to-leading
              order differential cross sections, and their matching to
              parton shower simulations}",
    eprint = "1405.0301",
    archivePrefix = "arXiv",
    primaryClass = "hep-ph",
    doi = "10.1007/JHEP07(2014)079",
    journal = "JHEP",
    volume = "07",
    pages = "079",
    year = "2014"
}

@article{Staub:2013tta,
    author = "Staub, Florian",
    title = "{SARAH 4: A tool for (not only SUSY) model builders}",
    eprint = "1309.7223",
    archivePrefix = "arXiv",
    primaryClass = "hep-ph",
    doi = "10.1016/j.cpc.2014.02.018",
    journal = "Comput. Phys. Commun.",
    volume = "185",
    pages = "1773--1790",
    year = "2014"
}

@article{Porod:2003um,
    author = "Porod, Werner",
    title = "{SPheno, a program for calculating supersymmetric spectra,
              SUSY particle decays and SUSY particle production at
              e+ e- colliders}",
    eprint = "hep-ph/0301101",
    archivePrefix = "arXiv",
    doi = "10.1016/S0010-4655(03)00222-4",
    journal = "Comput. Phys. Commun.",
    volume = "153",
    pages = "275--315",
    year = "2003"
}

@article{Porod:2011nf,
    author = "Porod, Werner and Staub, Florian",
    title = "{SPheno 3.1: extensions including flavour, CP-phases
              and models beyond the MSSM}",
    eprint = "1104.1573",
    archivePrefix = "arXiv",
    primaryClass = "hep-ph",
    doi = "10.1016/j.cpc.2012.05.021",
    journal = "Comput. Phys. Commun.",
    volume = "183",
    pages = "2458--2469",
    year = "2012"
}

@article{Sjostrand:2014zea,
    author = "Sjostrand, Torbjorn and Ask, Stefan and Christiansen, Jesper R. and Corke, Richard and Desai, Nishita and Ilten, Philip and Mrenna, Stephen and Prestel, Stefan and Rasmussen, Christine O. and Skands, Peter Z.",
    title = "{An Introduction to PYTHIA 8.2}",
    eprint = "1410.3012",
    archivePrefix = "arXiv",
    primaryClass = "hep-ph",
    doi = "10.1016/j.cpc.2015.01.024",
    journal = "Comput. Phys. Commun.",
    volume = "191",
    pages = "159--177",
    year = "2015"
}

@article{deFavereau:2013fsa,
    author = "de Favereau, J. and Delaere, C. and Demin, P. and Giammanco, A.
              and Lemaitre, V. and Mertens, A. and Selvaggi, M.",
    collaboration = "DELPHES 3",
    title = "{DELPHES 3, A modular framework for fast simulation of a generic collider experiment}",
    eprint = "1307.6346",
    archivePrefix = "arXiv",
    primaryClass = "hep-ex",
    doi = "10.1007/JHEP02(2014)057",
    journal = "JHEP",
    volume = "02",
    pages = "057",
    year = "2014"
}

@article{Cacciari:2008gp,
    author = "Cacciari, Matteo and Salam, Gavin P. and Soyez, Gregory",
    title = "{The anti-$k_t$ jet clustering algorithm}",
    eprint = "0802.1189",
    archivePrefix = "arXiv",
    primaryClass = "hep-ph",
    doi = "10.1088/1126-6708/2008/04/063",
    journal = "JHEP",
    volume = "04",
    pages = "063",
    year = "2008"
}

@article{Cowan:2010js,
    author = "Cowan, Glen and Cranmer, Kyle and Gross, Eilam and Vitells, Ofer",
    title = "{Asymptotic formulae for likelihood-based tests of new physics}",
    eprint = "1007.1727",
    archivePrefix = "arXiv",
    primaryClass = "physics.data-an",
    doi = "10.1140/epjc/s10052-011-1554-0",
    journal = "Eur. Phys. J. C",
    volume = "71",
    pages = "1554",
    year = "2011",
    note = "[Erratum: Eur. Phys. J. C 73, 2501 (2013)]"
}

@article{Krauss:2012ku,
    author = "Krauss, Manuel E. and O'Leary, Ben and Porod, Werner and Staub, Florian",
    title = "{Implications of gauge kinetic mixing on $Z'$ and slepton production at the LHC}",
    eprint = "1206.3513",
    archivePrefix = "arXiv",
    primaryClass = "hep-ph",
    doi = "10.1103/PhysRevD.86.055017",
    journal = "Phys. Rev. D",
    volume = "86",
    pages = "055017",
    year = "2012"
}

\end{document}